\documentclass[aps,reprint,amsmath,amssymb,groupaddress,superscriptaddress,prl]{revtex4-2}

\usepackage[utf8]{inputenc}
\usepackage{amsmath}
\usepackage{amsfonts}
\usepackage{amssymb}
\usepackage{bm}
\usepackage{color}
\usepackage{graphicx}
\usepackage{soul}
\usepackage{CJK}
\usepackage{xcolor}
\usepackage{mathtools}
\usepackage{lipsum}
\usepackage{comment}

\usepackage{xcolor}
\usepackage[colorlinks,citecolor=blue,linkcolor=red,anchorcolor=red,urlcolor=blue]{hyperref}

\begin{document}
\begin{CJK*}{UTF8}{gbsn}

\title{Engineering Ratchet-Based Particle Separation via Shortcuts to Isothermality}
\author{Xiu-Hua Zhao}
\affiliation{Department of Physics，Beijing Normal University, Beijing, 100875, China}
\author{Z. C. Tu}
\affiliation{Department of Physics，Beijing Normal University, Beijing, 100875, China}
\affiliation{Key Laboratory of Multiscale Spin Physics, Ministry of Education, China}
\author{Yu-Han Ma}
 \email{yhma@bnu.edu.cn}
\affiliation{Department of Physics，Beijing Normal University, Beijing, 100875, China}
\affiliation{Graduate School of China Academy of Engineering Physics, No. 10 Xibeiwang East Road, Haidian District, Beijing, 100193, China}

\begin{abstract}
Microscopic particle separation plays vital role in various scientific and industrial domains. In this Letter, we propose a universal non-equilibrium thermodynamic approach, employing the concept of Shortcuts to Isothermality, to realize controllable separation of overdamped Brownian particles. By utilizing a designed ratchet potential with temporal period $\tau$, we find in the slow-driving regime that the average particle velocity $\Bar{v}_s\propto\left(1-D/D^*\right)\tau^{-1}$, indicating that particles with different diffusion coefficients $D$ can be guided to move in distinct directions with a preset $D^*$. Furthermore, we reveal that there exists an extra energetic cost with a lower bound $W_{\rm{ex}}^{(\rm{min})}\propto\mathcal{L}^{2}\Bar{v}_s$, alongside a quasi-static work consumption. Here, $\mathcal{L}$ is the thermodynamic length of the driving loop in the parametric space. We numerically validate our theoretical findings and illustrate the optimal separation protocol (associated with $W_{\rm{ex}}^{(\rm{min})}$) with a sawtooth potential. This study establishes a bridge between thermodynamic process engineering and particle separation, paving the way for further explorations of thermodynamic constrains and optimal control in ratchet-based particle separation.

\end{abstract}

\maketitle
\end{CJK*}
\textit{Introduction}.---Particle separation, a fundamental process with broad applications in various scientific and industrial domain such as chemistry, biotechnology, materials science, environmental science, and food industry, has attracted significant research interest~\cite{harnisch_selectivity_2009,xie_critical_2014,sholl_seven_2016,makanyire_separation_2016,yang_lithium_2018,nouri_nanofluidic_2021,mei_bioinspired_2022}. Conventional separation methods relying on external forces or physical barriers inherently exhibit limitations in terms of efficiency, selectivity, and adaptability across diverse particle types~\cite{zamboulis_metal_2011,reguera_entropic_2012,zhang_selective_2020,yoon_review_2019,slapik_tunable_2019,marbach_active_2017,park_maximizing_2017,goh_membrane_2018,tang_ion_2020,epsztein_towards_2020}. For example, the membrane-based separation technology, extensively studied for water treatment and energy conversion, suffers from the fouling and instability issues~\cite{goh_membrane_2018,tang_ion_2020}. To overcome these limitations and achieve efficient separation applicable to a wider range of particle types, researchers are actively exploring innovative methods and techniques. 

Among the various of approaches explored, ratchet-based approach emerged as a highly promising avenue for particle separation~\cite{rousseletDirectionalMotionBrownian1994,faucheux_selection_1995,bader_dna_1999,matthias_asymmetric_2003}. By utilizing the driving force induced by spatially asymmetric potential, ratchet-based separation achieves directed motion of particles~\cite{rousseletDirectionalMotionBrownian1994,doering_randomly_1995,faucheux_selection_1995,parrondo_efficiency_1998,bader_dna_1999,reimann_brownian_2002,parrondo_energetics_2002,matthias_asymmetric_2003,lau_electron_2020,nicollier_nanometer-scale-resolution_2021,herman_ratchet-based_2023}, making it applicable for separating Brownian particles with different diffusion coefficients. So far, in the studies of ratchet-based particle separation, significant emphasis has been placed on particle current and its optimization~\cite{doering_randomly_1995,rozenbaum_high-temperature_2008,chr_germs_diffusion_2013,kedem_how_2017,kanada_diffusion_2018,herman_ratchet-based_2023}. However, existing theoretical results in certain limiting cases are too complex for further exploration~\cite{doering_randomly_1995,rozenbaum_high-temperature_2008,chr_germs_diffusion_2013,kanada_diffusion_2018}. Moreover, as a thermodynamic task, particle separation inevitably incurs an energetic cost, optimizing which is crucial for achieving energetically efficient separation~\cite{fane1987efficient,tsirlin2002finite,huang2010low,sholl_seven_2016,chen2023geodesic}. To the best of our knowledge, there is currently a dearth of theoretical studies investigating the energetic cost of generic ratchet-based particle separation. The primary challenge leading to these bottlenecks lies in analytically solving the particle's evolution in a periodically asymmetric potential. 

\begin{figure}[!tbh]
\includegraphics[width=1\columnwidth]{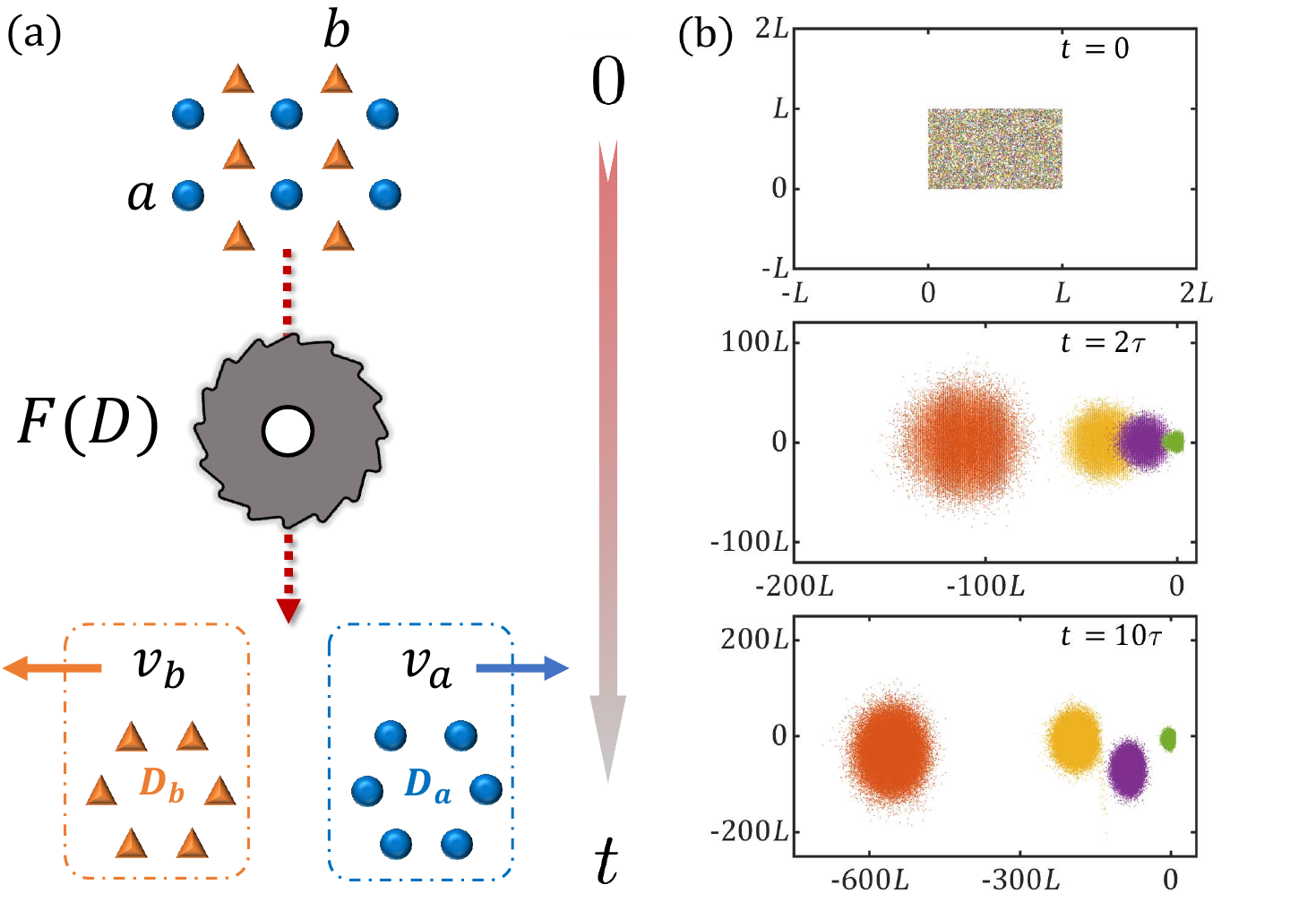}
\caption{(a) Schematic diagram of ratchet-based particle separation using the Shortcuts to Isothermality. With the designed potential, particles with different diffusion coefficients experience distinct driving forces, enabling directed movement in different directions. (b) Two-dimensional separation of four types of particles. $L$ and $\tau$ represent the spatial and temporal period of the ratchet potential, respectively.}
\label{fig: illustration}
\end{figure}
Recent advancements in thermodynamic process control~\cite{guery-odelin_shortcuts_2019,nakamura_fast-forward_2020,guery-odelin_driving_2023,li_shortcuts_2017,li_geodesic_2022,li_nonequilibrium_2023} offer new insight into this issue, wherein reverse engineering playing a crucial role in tackling the aforementioned challenge. In this Letter, we employ the Shortcuts to Isothermality (ScI)~\cite{li_shortcuts_2017} to revolutionize ratchet-based particle separation. ScI, a notable advancement in nonequilibrium thermodynamics, serves as a transformative tool by facilitating rapid evolution of systems to be controlled towards the thermal equilibrium state of their original Hamiltonians through the auxiliary potentials~\cite{li_shortcuts_2017,li_geodesic_2022}. Physically, particles experience an effective driving force~\cite{supplemental_material} that depends on their diffusion coefficient $D$, as depicted in Fig. \ref{fig: illustration}(a), leading to the separation of particles with varying $D$. Leveraging the remarkable capabilities offered by ScI, the designed particle evolution in ratchet-based separation is realized, allowing for tractable analytical discussion on typical thermodynamic quantities including particle flux and energy consumption. Moreover, our method can be extended to higher-dimensional cases, as illustrated in Fig. \ref{fig: illustration}(b), enabling efficient separation of various types of particles in distinct directions.

\textit{Framework}.---As the foundation of this study, we first incorporate ScI into the ratchet-based particle transport schemes. Consider a one-dimensional over-damped Brownian particle coupled to a bath at constant temperature $T$ and driven by a spatial periodic potential $U_o\equiv U_o(x,\Vec{\lambda})$ with period $L$. Here $\Vec{\lambda}\equiv\Vec{\lambda}(t)$ is a parametric vector with $N\ge 1$ components respectively dependent on time $t$. According to ScI, a designed auxiliary potential $U_a^*(x,t)$ satisfying~\cite{supplemental_material}
\begin{equation}
    \partial_x U_a^*(x,t)=\frac{1}{ \beta D^*}\frac{\Dot{\Vec{\lambda}}\cdot\left[\Vec{f}_o(x,\Vec{\lambda})+\Vec{j}(\Vec{\lambda})\right]}{\rho_o(x,\Vec{\lambda})}
    \label{eq-auxiliary-force}
\end{equation}
is exerted on the particle to make its evolution controllable. Here, $D^*$ serves as a reference diffusion coefficient but does not necessarily need to be identical to $D$ of the Brownian particle, which is different from the original ScI theory~\cite{li_shortcuts_2017}. $\rho_o(x,\Vec{\lambda})\equiv e^{-\beta U_o(x,\Vec{\lambda})}/\int_0^L e^{-\beta U_o(x,\Vec{\lambda})}dx$ is the normalized equilibrium probability density over one period $0\le x\le L$ with $\beta\equiv1/(k_{\mathrm{B}}T)$ and $k_{\mathrm{B}}$ being the Boltzmann constant, $\Vec{f}_o(x,\Vec{\lambda})\equiv\int_0^x \Vec{\nabla}_{\lambda}\rho_o(x^{\prime},\Vec{\lambda}) d x^{\prime}$, and $\Vec{j}(\Vec{\lambda})$ is an arbitrary $N$-dimensional vector function.  $\Vec{\nabla}_{\lambda}\equiv (\partial_{\lambda_1}, \partial_{\lambda_2}, \dots, \partial_{\lambda_N})$ and $\mathcal{\dot{O}}$ represents the time derivative of an arbitrary physical quantity
$\mathcal{O}$. 

Associated with the total potential $U^*(x,t)=U_o(x,\Vec{\lambda})+U_a^*(x,t)$, the evolution of the probability density $\rho(x,t)$ of the particle is governed by the over-damped Fokker-Planck equation~\cite{reichl2016modern}
\begin{equation}
    \partial_t \rho(x,t)=-\partial_x\hat{J}_t \rho(x,t),\label{eq-FPeq}
\end{equation}
where $\hat{J}_t\equiv -D[\beta\partial_x U^*(x,t)+\partial_x]$ is the current operator. Due to the periodicity of the current operator, it suffices to solve Eq.~(\ref{eq-FPeq}) in one period $x\in [0,L]$~\cite{reimann_brownian_2002}. Specifically, we define the reduced probability density $\rho_s(x,t)\equiv\sum_{n}\rho(x+nL,t)$ and the reduced probability current $J_s(x,t)\equiv\sum_{n}J(x+nL,t)$, where $n\in \mathbb{Z}$ and the probability current reads
\begin{equation}
    J(x,t)= \hat{J}_t\rho(x,t)\label{eq-current-def}.
\end{equation}
Providing that $\rho(x,t)$ is a normalized solution of Eq.~(\ref{eq-FPeq}), $\rho_s(x,t)$ is also a solution which satisfies the periodic condition $\rho_s(x,t)=\rho_s(x+L,t)$ as well as the conservation condition $\int_0^L\rho_s(x,t)dx=1$. Therefore, the relation between $\rho_s(x,t)$ and $J_s(x,t)$ is the same as Eq.~(\ref{eq-current-def}), and the ensemble-averaged velocity of the Brownian particle is defined as $v_s(t)\equiv\int_0^L J_s(x,t)dx$.

When all the Brownian particles of interest possess the same diffusion coefficient, it is natural to set $D^*=D$. As the result of the original ScI theory, we obtain $\rho_s(x,t)=\rho_o(x,\Vec{\lambda})$ from Eq.~(\ref{eq-FPeq}) and the initial condition $\rho_s(x,0)=\rho_o(x,0)$. For cases involving multiple particle ensembles with different diffusion coefficients, we need to solve $\rho_s(x,t)$ at $D^*\neq D$. To carry out further analytical discussion, we assume that the parametric vector $\Vec{\lambda}$ changes slowly enough over time, so that $\rho_s(x,t)$ can be expanded up to the linear term of $\Dot{\Vec{\lambda}}$~\cite{cavina2017slow,ma2022minimal} as follows
\begin{equation}
    \rho_s(x,t)\approx \rho_o(x,\Vec{\lambda})+\Dot{\Vec{\lambda}}\cdot\Vec{\psi}(x,t),\label{eq-distri-expre}
\end{equation}
where $\Vec{\psi}(x,t)$ is a $N$-dimensional vector function to be solved. The equilibrium state is recovered when $\Dot{\Vec{\lambda}}=0$. Substituting Eq.~(\ref{eq-distri-expre}) into Eq.~(\ref{eq-FPeq}) and neglecting the terms containing quadratic time derivative, we obtain
\begin{equation}
    \frac{\partial\Vec{\psi}(x,t)}{\partial x}+\beta\frac{\partial U_o}{\partial x}\Vec{\psi}(x,t)=\left(\frac{1}{D}-\frac{1}{D^*}\right)\left[\Vec{f}_o(x,\Vec{\lambda})+\Vec{C}(t)\right]
    \label{eq-psi-eq}
\end{equation}
with $\Vec{C}(t)$ the constant of integration. Solving Eq.~(\ref{eq-psi-eq}) with boundary conditions $\Vec{\psi}(0,t)=\Vec{\psi}(L,t)$ and $\int_0^L\Vec{\psi}(x,t)dx=0$, we find $\Vec{C}(t)=-\langle\Vec{f}_o(x,\Vec{\lambda})\rangle_+$, where $\langle\cdots\rangle_{\pm}\equiv\int_0^L e^{\pm\beta U_o(x,\Vec{\lambda})}\cdots dx/Z_{\pm}(\Vec{\lambda})$ and $Z_{\pm}(\Vec{\lambda})\equiv\int_0^Le^{\pm \beta U_o(x,\Vec{\lambda})}dx$. $\Vec{\psi}(x,t)=\Vec{\psi}(x,\Vec{\lambda})$ is given in~\cite{supplemental_material}. It follows from Eqs.~(\ref{eq-current-def}) and (\ref{eq-psi-eq}) that
\begin{equation}
\label{Js}
    J_s(x,t)=-\Dot{\Vec{\lambda}}\cdot\Vec{f}(x,\Vec{\lambda})+\left(1-\frac{D}{D^*}\right)\Dot{\Vec{\lambda}}\cdot\langle \Vec{f}(x,\Vec{\lambda})\rangle_+,
\end{equation}
where $\Vec{f}(x,\Vec{\lambda})\equiv \Vec{f}_o(x,\Vec{\lambda})+\Vec{j}(\Vec{\lambda})$.

\textit{Separating particles with different $D$}.---We then investigate the particle flux in steady state. To induce steady-state evolution, we consider that the Brownian particles are periodically driven, namely, $\Vec{\lambda}(t)=\Vec{\lambda}(t+\tau)$ with the temporal period $\tau$. After enough periods, $\rho_s(x,t)$ will enter steady periodic evolution independent of the initial condition~\cite{supplemental_material}. 
According to Eq.~\eqref{Js}, the average reduced probability current over a temporal period $\Bar{J}_s\equiv \tau^{-1}\int_{t_0}^{t_0+\tau}J_s(x,t)dt$ is specifically obtained as
\begin{equation}
        \Bar{J}_s=\frac{1}{\tau}\left(1-\frac{D}{D^*}\right)\Phi_{\mathrm{rev}}-\frac{1}{\tau}\frac{D}{D^*}\oint_I d\Vec{\lambda}\cdot\Vec{j}(\Vec{\lambda}),
        \label{eq-Jbar-expre}
\end{equation}
where 
\begin{equation}
    \Phi_{\mathrm{rev}}\equiv\oint_I d\Vec{\lambda}\cdot\frac{\int_0^L e^{\beta U_o(x,\Vec{\lambda})}\Vec{\nabla}_{\lambda}\int_0^x\rho_o(x^{\prime},\Vec{\lambda})dx^{\prime}dx}{\int_0^L e^{\beta U_o(x,\Vec{\lambda})}dx}\label{eq-phirev}
\end{equation}
is the integrated flow of reversible ratchets~\cite{parrondo_efficiency_1998,parrondo_energetics_2002} and $I$ is a closed trajectory of $\Vec{\lambda}$. 

As one of the main results of this Letter, Eq.~(\ref{eq-Jbar-expre}) allows Brownian particles with different $D$ to averagely move at different velocities, thereby enabling their spatial separation. We stress here that i) a spatially asymmetric $U_o(x,\Vec{\lambda})$ is necessary for generating non-zero $\Phi_{\text{rev}}$ \cite{supplemental_material}; ii) since $\Phi_{\text{rev}}$ and $\Bar{J}_s$ are geometric quantities in the $N$-dimensional parametric space that only depend on the geometry of $I$, $N\ge2$ is required to result in non-zero $\Phi_{\text{rev}}$ and $\Bar{J}_s$. Obviously, the velocity difference among different particles can be accordingly changed with $\Vec{j}(\Vec{\lambda})$, $\Phi_{\mathrm{rev}}$, $\tau$, and the setting of $D^*$. In particular, when $\Vec{j}(\Vec{\lambda})=0$, the particles with $D>D^*$ and those with $D<D^*$ will move in opposite directions, which is consistent with a recent numerical study \cite{herman_ratchet-based_2023}. In real-world circumstances, different types of particles possess different $D$ due to variations in their shape, size, surface structure, and other characteristics~\cite{munRoleSpecificInteractions2014,chanEffectsShapesSolute2015,iwahashiEffectsMolecularSize2007,marreroGaseousDiffusionCoefficients2009}. Therefore, by choosing an appropriate $D^*$ to design $U_a^*(x,t)$ according to Eq.~\eqref{eq-auxiliary-force}, the desired particle separation can be achieved. Noticing $\Bar{J}_s$ is independent of $x$, it is naturally for us to define the time-ensemble-averaged velocity of the particles as $\Bar{v}_s\equiv \tau^{-1}\int_{t_0}^{t_0+\tau}v_s(t)dt=\Bar{J}_sL$. For practical case with $D^*\sim10^{-5}{\rm{cm^2/s}},L\sim0.1{\rm{\mu m}},f=\tau^{-1}\sim100{\rm{kHz}}$~\cite{herman_ratchet-based_2023}, according to Eq.~\eqref{eq-Jbar-expre}, $\Delta D /D^*\sim1\%$ can result in a velocity difference $\Delta \Bar{v}_s\sim0.1\rm{mm/s}$.

\textit{Energetic cost for particle separation}.---As a thermodynamic task, the energy consumption in driving the particle is another typical quantity that requires significant attention, which can be analyzed with stochastic thermodynamics~\cite{sekimotoStochasticEnergetics2010,seifertStochasticThermodynamicsFluctuation2012}. When the particles of interest have entered the steady periodic state, their energetics may be captured through the above-solved reduced probability density and reduced probability current. 
According to the 1st law of thermodynamics, the ensemble-averaged work needed in driving the particle is $W=\Delta E-Q$, where $Q$ is the ensemble-averaged heat absorbed by the particle and $\Delta E$ is the variation of internal energy $E(t)\equiv\int U^*(x,t)\rho(x,t)dx$. The total potential $U^*(x,t)$ obtained by adding the integral of Eq.~(\ref{eq-auxiliary-force}) on $U_o(x,\Vec{\lambda})$ is a tilted ratchet potential
\begin{equation}
    U^*(x,t)=V^*(x,t)+\frac{\varepsilon^*(t)}{L}x,\label{eq-total-potential}
\end{equation}
where
\begin{equation}
    \varepsilon^*(t)=\frac{1}{\beta D^*}Z_{-}(\Vec{\lambda})Z_{+}(\Vec{\lambda})\Dot{\Vec{\lambda}}\cdot\langle \Vec{f}(x,\Vec{\lambda})\rangle_{+}
\end{equation}
is the variation of $U_a^*(x,t)$ from $x_0$ to $x_0+L$, and  $V^*(x,t)\equiv U_o(x,\Vec{\lambda})+\int_0^x\partial_{x^{\prime}}U_a^*(x^{\prime},t)dx^{\prime}-L^{-1}\varepsilon^*(t)x$ is a spatial periodic function with period $L$. Here, we have set $U_a^*(0,t)=0$. Then the internal energy of the particle turns out to be~\cite{supplemental_material}
\begin{equation}
    E(t)=\int_0^L V^*(x,t)\rho_s(x,t)dx+\frac{\varepsilon^*(t)}{L}\langle x\rangle
    \label{eq-internal-energy}
\end{equation}
with $\langle x \rangle\equiv \int_0^L x\rho_s(x,t)dx$ being the ensemble-averaged position of the particle. For periodic driving with $\Vec{\lambda}(t_0)=\Vec{\lambda}({t_0+\tau})$ and $\Dot{\Vec{\lambda}}(t_0)=\Dot{\Vec{\lambda}}({t_0+\tau})$, the variation of the first term in Eq.~(\ref{eq-internal-energy}) vanishes in a temporal period. By noticing $\varepsilon^*(t_0)=\varepsilon^*(t_0+\tau)$ and $\Delta \langle x \rangle=\Bar{v}_s \tau$, we obtain the internal energy variation as
\begin{equation}
\Delta E= \varepsilon^*(t_0)\Bar{J}_s\tau=\left(1-\frac{D}{D^*}\right)\varepsilon^*(t_0)\Phi_{\mathrm{rev}},
\end{equation}
which depends on the initial condition of the driving protocol. Such a initial value dependence diminishes as the particle transport duration increases~\cite{supplemental_material}. Hereafter, unless otherwise stated, we take $\Vec{j}(\Vec{\lambda})=0$ for simplicity. 


Furthermore, the heat current reads $\Dot{Q}(t)\equiv\int U^*(x,t)\partial_t\rho(x,t)dx$~\cite{sekimotoStochasticEnergetics2010,seifertStochasticThermodynamicsFluctuation2012}, according to which the heat absorption in a temporal period is given as~\cite{supplemental_material}

\begin{equation}
    Q=-\int_{t_0}^{t_0+\tau}dt\Dot{\lambda}_{\alpha}\Dot{\lambda}_{\beta}G_{\alpha\beta}(\Vec{\lambda}).\label{eq-heat}
\end{equation}
Here,
\begin{equation}
\begin{aligned}
    G_{\alpha\beta}(\Vec{\lambda})\equiv&\frac{1}{\beta D^*}Z_{-}(\Vec{\lambda})Z_{+}(\Vec{\lambda})\bigg[\langle f_{\alpha}(x,\Vec{\lambda})f_{\beta}(x,\Vec{\lambda})\rangle_{+}\\
    &-\left(1-\frac{D}{D^*}\right)\langle f_{\alpha}(x,\Vec{\lambda})\rangle_{+}\langle f_{\beta}(x,\Vec{\lambda})\rangle_{+}\bigg]
\end{aligned}\label{metric}
\end{equation}
 is a positive semi-definite matrix \cite{supplemental_material} with $\alpha,\beta=1,2,\dots,N$, and the Einstein notation has been adopted hereafter.

For a given closed driving trajectory in the parametric space, Cauchy-Schwarz inequality implies that the heat release in Eq.~(\ref{eq-heat}) is bounded from below as $-Q\ge\mathcal{L}^2/\tau$, where $\mathcal{L}\equiv\int_{0}^{\tau}dt\sqrt{\Dot{\lambda}_{\alpha}\Dot{\lambda}_{\beta}G_{\alpha\beta}(\Vec{\lambda})}$ is the so-called thermodynamic length~\cite{salamonThermodynamicLengthDissipated1983,crooksMeasuringThermodynamicLength2007,sivak2012thermodynamic,li_geodesic_2022,ma2022minimal} of the driving loop. Therefore, we have the work cost satisfying $W\ge\Delta E+\mathcal{L}^2/\tau$, which, together with Eq. \eqref{eq-Jbar-expre}, yields the second main result of this Letter
\begin{equation}
W\ge\Delta E+\frac{D^*\mathcal{L}^2\Bar{J}_s}{\left(D^*-D\right)\Phi_{\mathrm{rev}}},
\label{eq-Workineq}
\end{equation}
where the equal sign is saturated when the integrand of $\mathcal{L}$ is a time-independent constant~\cite{crooksMeasuringThermodynamicLength2007,sivak2012thermodynamic}, namely,  
\begin{equation}
\tau\sqrt{\Dot{\lambda}_{\alpha}\Dot{\lambda}_{\beta}G_{\alpha\beta}(\Vec{\lambda})}-\mathcal{L}=0.
\label{eq-optimal}
\end{equation}
Equation \eqref{eq-Workineq} indicates that the minimal extra energetic cost ($W_{\rm{ex}}\equiv W-\Delta 
 E$ which is exactly the heat dissipation) for particle separation is directly proportional to the particle flow, namely, faster particle separation (shorter $\tau$) requires more work consumption for given $\Delta E$. In the quasi-static limit ($\Bar{J}_s\rightarrow0$), one has   $W\rightarrow\Delta E$. Moreover, for given parametric loop and $\tau$ (which correspond to a certain average particle velocity), Eq. \eqref{eq-optimal} determines the optimal driving protocol associated with the minimal $W_{\rm{ex}}$. 

\begin{figure}[!tbh]
    \includegraphics{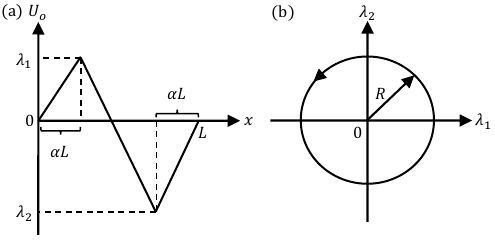}
    \includegraphics{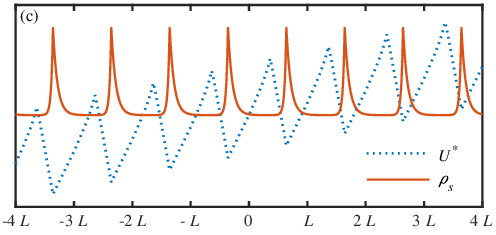}
    \caption{(a) One period of the original periodic potential $U_o(x,\Vec{\lambda})$. (b) The trajectory of the driving protocol $\Vec{\lambda}(t)$ is a anticlockwise circle with radius $R$. (c) The tilted total potential $U^*(x,t)$ and the periodic steady reduced probability density $\rho_s(x,t)$ at a certain time.}
    \label{fig: potential-protocol}
\end{figure}

\textit{Example}.---We illustrate our general theoretical framework with an example, where $U_o$ is specific as the sawtooth potential which, shown in Fig.~\ref{fig: potential-protocol}(a), is one of the most common ratchet potential. The height $\lambda_1$ and depth $\lambda_2$ of the potential serve as time-dependent parameters, i.e., $\Vec{\lambda}=(\lambda_1,\lambda_2)$. According to Eq.~(\ref{eq-Jbar-expre}), $U_o(x,\Vec{\lambda})$ relates to the average particle probability current via $\Phi_{\text{rev}}$. Hence, the shape of the potential as well as the geometry of the driving loop in the parametric space can be optimized to induce large particle current. After comprehensive evaluations~\cite{supplemental_material}, we find that it is favorable to set $\alpha=0.36$ and the driving loop as a circle in Fig.~\ref{fig: potential-protocol}(b). Fig.~\ref{fig: potential-protocol}(c) is a snapshot of the total potential $U^*(x,t)$ and the steady reduced probability density $\rho_s(x,t)$ respectively according to Eq.~(\ref{eq-total-potential}) and Eq.~(\ref{eq-distri-expre}). $\rho_s(x,t)$ and the gradient of $U^*(x,t)$ are both periodic in infinite space. The dynamic equation of the over-damped Brownian particles reads $\Dot{x}=-\beta D\partial_xU^*(x,t)+\sqrt{2D}\xi(t)$~\cite{zwanzigNonequilibriumStatisticalMechanics2001}, where the normalized Gaussian white noise $\xi(t)$ satisfies $\langle\xi(t)\rangle=0$ and $\langle\xi(t)\xi(t^{\prime})\rangle=\delta(t-t^{\prime})$. We simulate the movement of $10^5$ particles by solving this equation with Euler algorithm \cite{supplemental_material}. The quantities in the simulation are nondimensionalized by $L$, $\beta^{-1}$ and $D^*$. $\mathcal{\Tilde{O}}$ denotes the dimensionless $\mathcal{O}$. For example, $\widetilde{v_s}\equiv D^{*-1}Lv_s$.

\begin{figure}[!tbh]
\includegraphics{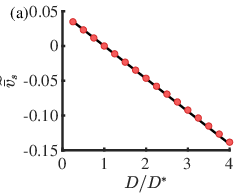}
\includegraphics{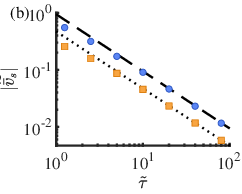}
\includegraphics{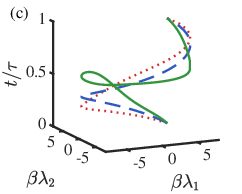}
\includegraphics{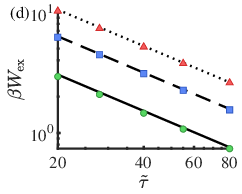}
\caption{(a) The time-ensemble-averaged velocity $\widetilde{\Bar{v}_s}$ as a function of $D/D^*$ with $\Tilde{\tau}=20$. (b) $\widetilde{\Bar{v}_s}$ as a function of $\Tilde{\tau}$ for $D/D^*=2$ (circles and dashed line) and $D/D^*=1/2$ (squares and dotted line). (c) Three different driving protocols, i.e., path-I (red dotted line), path-II (blue dashed line), and path-III (green solid line). (d) The extra energetic cost $W_{\rm{ex}}$ as a function of $\Tilde{\tau}$ with $D/D^*=2$. The three data series from top to bottom respectively correspond to path-I, path-II and path-III. The marks and lines in (a)(b)(d) are simulation results and analytical predictions, respectively. In the simulations, characteristic quantities are set as $L=1$, $\beta=1$, $D^*=1$, and $\beta R=7$, the number of particles is $N=10^5$ and the time step is $\Delta\Tilde{t}=10^{-4}$.}
\label{fig-verify}
\end{figure}

We first validate the effectiveness of Eq.~(\ref{eq-Jbar-expre}). The time-ensemble-averaged velocities $\widetilde{\Bar{v}_s}$ of particles for different diffusion coefficients are plotted in Fig.~\ref{fig-verify}(a) with $\Tilde{\tau}\equiv D^*L^{-2}\tau=20$. The simulation data (red circles) are in good alignment with the theoretical prediction (solid line). 
In Fig.~\ref{fig-verify}(b), we illustrate $\widetilde{\Bar{v}_s}$ as a function of  $\Tilde{\tau}$ for $D/D^*=2$ (circles) and $D/D^*=1/2$ (squares). As expected, the simulated marks coincide well with the theoretical lines in the slow-driving regime ($\Tilde{\tau} \gg 1$). Furthermore, the energetics of the particle can be consistently obtained in simulations. By definition~\cite{sekimotoStochasticEnergetics2010}, the absorbed heat and the input work of a particle from $t$ to $t+\Delta t$ are respectively $\Delta q=U^*(x+\Delta x,t+\Delta t)-U^*(x,t+\Delta t)$ and $\Delta w=U^*(x,t+\Delta t)-U^*(x,t)$, where $\Delta x$ is the position variation within $\Delta t$. We now test Eq. \eqref{eq-Workineq} with three different protocols $f(s)$ associated with the driving loop (Fig. \ref{fig: potential-protocol}(b)) $\lambda_1=R\cos[2\pi f(s)+\theta_{0}]$, $\lambda_2=R\sin[2\pi f(s)+\theta_{0}]$, where $s\equiv t/\tau$ and $\theta_{0}=\pi/6$. The time-dependent paths are demonstrated in Fig.~\ref{fig-verify}(c): Path-I: $f_1(s)=\sum_{i=1}^{i=3}a_{i}s^{i}$ with $a_1=2,a_2=-3,a_3=2$ (red dotted line) \cite{protocol}, Path-II: $f_2(s)=s$ (blue dashed line), and Path-III: the optimal protocol (green solid line) obtained numerically from Eq.~\eqref{eq-optimal}~\cite{supplemental_material}. The corresponding extra work $W_{\rm{ex}}$ are illustrated in Fig.~\ref{fig-verify}(d), where the dotted and dashed lines (plotted with Eq.~(\ref{eq-heat})) and the solid line (the lower bound of Eq.~(\ref{eq-Workineq})) agree well with the numerical results (marks). Clearly, the optimal protocol~\cite{optimal} indeed lead to lower $W_{\rm{ex}}$ (green circles) than those associated with Path-I (red triangles) and Path-II (blue squares). 

Finally, we would like to make three remarks here. First, although generated under path-I, the results in Fig.~\ref{fig-verify}(a) and (b) are independent of the specific choice of driving protocol $f(s)$ since the time-ensemble-averaged velocity is a geometric quantity (dynamic-independent) in the parametric space. Second, in the slow-driving regime (also known as long-time regime), the $1/\tau$-scaling exhibited by the particle flux (Fig.~\ref{fig-verify}(b)) and energetic cost (Fig.~\ref{fig-verify}(d)) is a typical manifestation of finite-time irreversibility~\cite{martinez2016brownian,li_shortcuts_2017,ma2020experimental,yuan2022Optimizing,li_geodesic_2022,ma2022minimal}. Third, for practical purposes, the developed ScI-ratchet can be straightforwardly generalized to higher-dimensional space to simultaneously separate more kinds of particles, as demonstrated in Fig.~\ref{fig: illustration}(b). Detailed information of this case is given in \cite{supplemental_material}.

\textit{Summary}.---
We develop a universal framework that integrates thermodynamic process engineering into ratchet-based particle transport, enabling directional separation of Brownian particles with different diffusion coefficients. By utilizing ScI, which allows controlled evolution of particles towards an thermal equilibrium distribution, we provide analytical expressions for key quantities in the particle separation process, such as particle flux, heat dissipation, and required work. With the thermodynamic geometry theory, we determine the optimal separation protocol that minimize the energetic consumption while maintaining a given particle flux. We also demonstrate the extensibility of this approach in higher-dimensional space. 

Currently, the combination of thermodynamic geometry and thermodynamic process control in optimizing practical thermodynamic tasks, such as heat engine optimization \cite{ma2018optimal,frim2022geometric,zhao2022low} and information erasure \cite{ma2022minimal,rolandi2023Collective}, has attracted widespread research interest. Our work provides new application scenarios for this area and lays the foundation for further incorporation of thermodynamic process engineering into ratchet-based particle separation. In relation to this, it is interesting to investigate ScI-assisted chiral separation~\cite{marcosSeparationMicroscaleChiral2009,bechingerActiveParticlesComplex2016,xuSortingChiralActive2022} and mass separation~\cite{slapik_tunable_2019,khatriMassSeparationAsymmetric2021,khatriInertialEffectsRectification2023,guptaSeparationInteractingActive2023}. Our proposed Brownian particle separation method and the corresponding theoretical predictions can be experimental realized and tested in some state-of-art platforms~\cite{albayThermodynamicCostShortcutstoisothermal2019,albayRealizationFiniterateIsothermal2020}. 

\textit{Acknowledgments}.---
Y. H. Ma thanks the National Natural Science Foundation of China for support under grant No. 12305037 and the Fundamental Research Funds for the Central Universities under grant No. 2023NTST017. Z. C. Tu thanks the National Natural Science Foundation of China for support under grant No. 11975050.
        
\bibliography{refs}

\begin{thebibliography}{73}
\bibitem [{protocol()}]{protocol}%
  \BibitemOpen
  \bibinfo {note} {In fact, only a free parameter exsits in this protocol,
  since $a_1+a_2+a_3=1$ and $2a_2+3a_3=0$ are imposed to ensure the continuity
  of $f(s)$ and its first derivative along the driving loop.}\BibitemShut
  {Stop}%
\bibitem [{optimal()}]{optimal}%
  \BibitemOpen
  \bibinfo {note} {The presented optimal path associated with the minimum
  $W_{\rm{ex}}$ on a given trajectory in the parametric space. There may exist
  different trajectories leading to the same particle flux but have different
  energy cost. Determining the optimal trajectory and the corresponding driving
  protocol is an open question for achieving globally optimal control in both
  geometric and dynamic regimes.}\BibitemShut {Stop}%
\end{thebibliography}
\end{document}


\begin{CJK*}{UTF8}{gbsn}

\title{Supplemental Materials for: \\ ``Engineering Ratchet-Based Particle Separation via Shortcuts to Isothermality''}
         
\author{Xiu-Hua Zhao}
\affiliation{Department of Physics，Beijing Normal University, Beijing, 100875, China}
\author{Z. C. Tu}
\affiliation{Department of Physics，Beijing Normal University, Beijing, 100875, China}
\affiliation{Key Laboratory of Multiscale Spin Physics, Ministry of Education, China}
\author{Yu-Han Ma}
 \email{yhma@bnu.edu.cn}
\affiliation{Department of Physics，Beijing Normal University, Beijing, 100875, China}
\affiliation{Graduate School of China Academy of Engineering Physics, No. 10 Xibeiwang East Road, Haidian District, Beijing, 100193, China}

\maketitle
\end{CJK*}
\onecolumngrid
This document is devoted to providing the detailed derivations and the supporting discussions to the main text of the Letter. The contents of the supplemental materials are listed as follows
\tableofcontents

\section{Framework}

\subsection{Auxiliary potential}

According to the Shortcuts to Isothermality (ScI) theory~\cite{li_shortcuts_2017}, an auxiliary potential $U_a$ is exerted on the Brownian particle originally driven by the time-dependent potential $U_o$, so that the probability distribution of the particle in the phase space is described by the equilibrium probability density of $U_o$ at constant temperature $T$. The form of $U_a$ is determined by the Fokker-Planck equation associated with the total potential $U_o+U_a$. Specifically, the Fokker-Planck equation for the one-dimensional over-damped Brownian particle is
\begin{equation}
    \frac{\partial\rho(x,t)}{\partial t}=D\frac{\partial}{\partial x}\left(\beta\frac{\partial U_o(x,\Vec{\lambda})}{\partial x}+\beta\frac{\partial U_a(x,t)}{\partial x}+\frac{\partial}{\partial x}\right)\rho(x,t),
\end{equation}
where $D$ is the diffusion coefficient of the particle and $\beta\equiv 1/(k_{\mathrm{B}}T)$. By substituting the target density $\rho(x,t)=\rho_o(x,\Vec{\lambda})\equiv e^{-\beta U_o(x,\Vec{\lambda})}/\int e^{-\beta U_o(x,\Vec{\lambda})}dx$ into the above equation, the auxiliary potential is found to satisfy~\cite{li_shortcuts_2017,li_geodesic_2022}
\begin{equation}
    \frac{\partial U_a(x,t)}{\partial x}=\frac{1}{\beta D}\frac{\Dot{\Vec{\lambda}}\cdot\left[\int_0^x\Vec{\nabla}_{\lambda}\rho_o(x^{\prime},\Vec{\lambda})dx^{\prime}+\Vec{j}(\Vec{\lambda})\right]}{\rho_o(x,\Vec{\lambda})}.
\end{equation}
To incorporate ScI into the ratchet-based particle transport scheme, considering that there are maybe particles with different diffusion coefficients under the same driving condition, we modify the above equation as follows
\begin{equation}
    \frac{\partial U_a^*(x,t)}{\partial x}=\frac{1}{\beta D^*}\frac{\Dot{\Vec{\lambda}}\cdot\left[\int_0^x\Vec{\nabla}_{\lambda}\rho_o(x^{\prime},\Vec{\lambda})dx^{\prime}+\Vec{j}(\Vec{\lambda})\right]}{\rho_o(x,\Vec{\lambda})},
\end{equation}
where $D^*$ servers as a reference diffusion coefficient and $\rho_o(x,\Vec{\lambda})$ is normalized on $x\in [0,L]$ since $U_o(x,\Vec{\lambda})$ is spatially periodic. The probability density of the particle with $D$ is no longer described by $\rho_o(x,\Vec{\lambda})$ unless $D=D^*$ or $\Dot{\Vec{\lambda}}=0$.

\subsection{Reduced probability density}

Considering the slow-driving regime, the reduced probability density is expanded as $\rho_s(x,t)=\rho_o(x,\Vec{\lambda})+\Dot{\Vec{\lambda}}\cdot\Vec{\psi}(x,t)$ where $\Vec{\psi}(x,t)$ is determined by Eq.~(5) of the main text. The general solution of Eq.~(5) is
\begin{equation}
    \Vec{\psi}(x,t)=\left(\frac{1}{D}-\frac{1}{D^*}\right)e^{-\beta U_o(x,\Vec{\lambda})}\int_0^x e^{\beta U_o(x^{\prime},\Vec{\lambda})}\left[\Vec{f}_o(x^{\prime},\Vec{\lambda})+\Vec{C}(t)\right]dx^{\prime}+\left(\frac{1}{D}-\frac{1}{D^*}\right)\Vec{B}(t)e^{-\beta U_o(x,\Vec{\lambda})}.\label{eq-psi-expr}
\end{equation}
From the conditions $\rho_{s}(0,t)=\rho_{s}(L,t)$, $\int_0^L\rho_{s}(x,t)dx=1$ and the definition of $\rho_o(x,\Vec{\lambda})$, we obtain $\Vec{\psi}(0,t)=\Vec{\psi}(L,t)$ and $\int_0^L\Vec{\psi}(x,t)dx=0$ which are used to determine the expressions of $\Vec{C}(t)$ and $\Vec{B}(t)$:
\begin{align}
    \Vec{C}(t)=&\Vec{C}(\Vec{\lambda})=-\frac{\int_0^L e^{\beta U_o(x,\Vec{\lambda})}\Vec{f}_o(x,\Vec{\lambda})dx}{\int_0^L e^{\beta U_o(x,\Vec{\lambda})}dx}\label{eq-C-expr}\\
    \Vec{B}(t)=&\Vec{B}(\Vec{\lambda})=-\frac{\int_0^L e^{-\beta U_o(x,\Vec{\lambda})}\int_0^x e^{\beta U_o(x^{\prime},\Vec{\lambda})}\left[\Vec{f}_o(x^{\prime},\Vec{\lambda})+\Vec{C}(\Vec{\lambda})\right]dx^{\prime}dx}{\int_0^L e^{-\beta U_o(x,\Vec{\lambda})}dx}.\label{eq-B-expr}
\end{align}
The expression of $\Vec{\psi}(x,t)=\Vec{\psi}(x,\Vec{\lambda})$ is totally determined by $U_o(x,\Vec{\lambda})$.

\begin{figure}
    \centering
\includegraphics{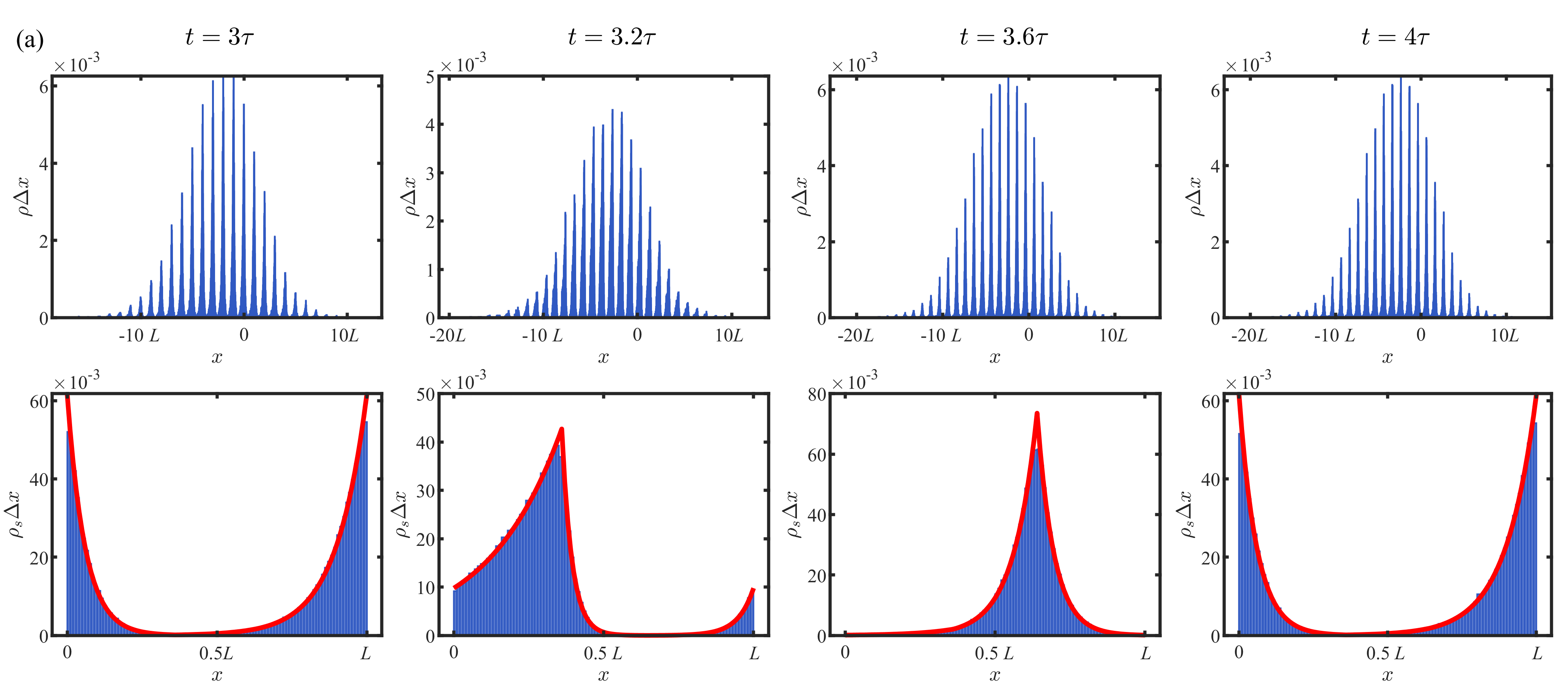}
\includegraphics{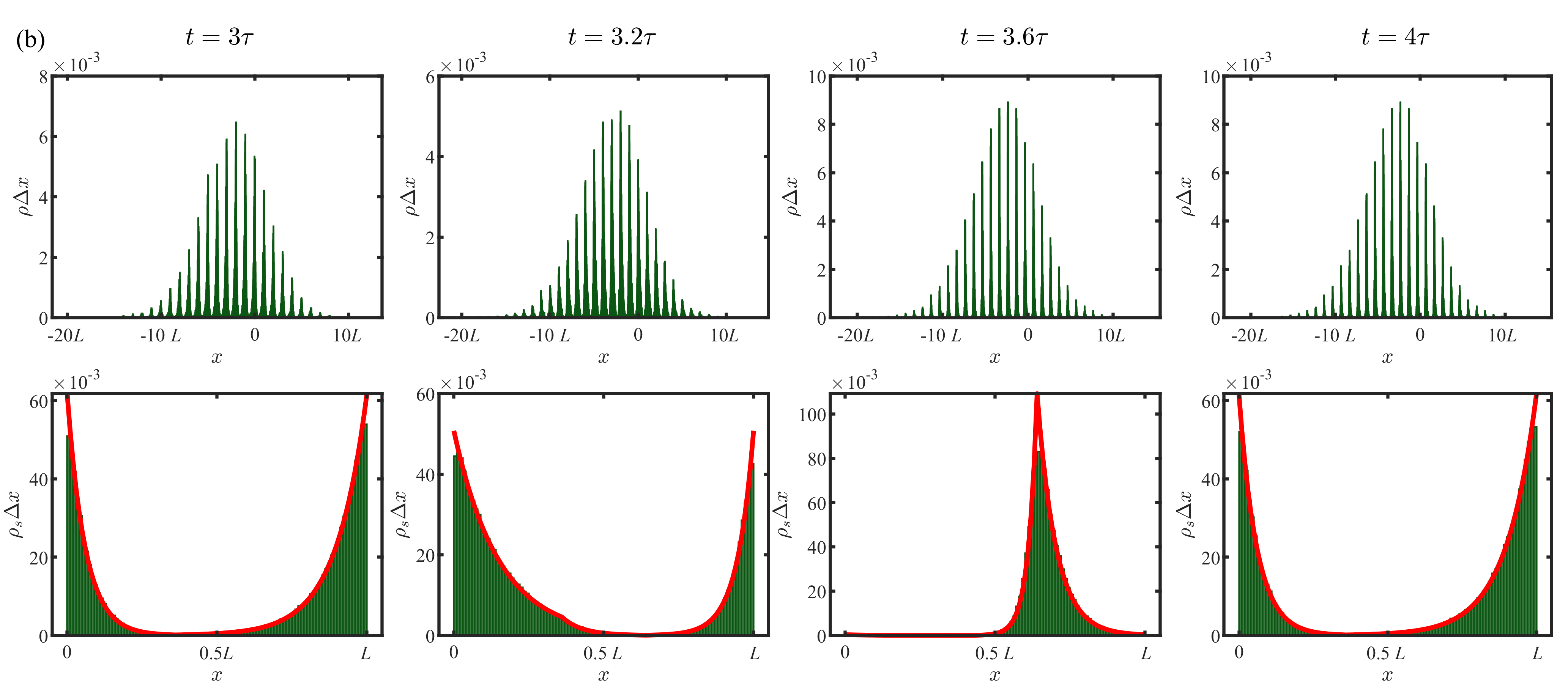}
    \caption{Evolution of the steady-state distribution. The histograms in the upper and lower panels represent actual and reduced distributions, respectively. The red solid lines are analytical predictions. (a) The distributions under the protocol along path-II. (b) The distributions under the protocol along path-III. The parameters in the simulations are $\Tilde{\tau}=20$, $D/D^*=2$, $\beta=L=D^*=1$, the particle number $N=10^5$ and the time step $\Delta\Tilde{t}=10^{-4}$. The width of the histogram bars is $\Delta x=10^{-2}L$.}
    \label{fig: steady-state}
\end{figure}

Next, we numerically demonstrate that the approximate analytical probability density $\rho_s(x,t)=\rho_o(x,\Vec{\lambda})+\Dot{\Vec{\lambda}}\cdot\Vec{\psi}(x,\Vec{\lambda})$ is consistent with the reduced distribution of the Brownian particles when they enter steady periodic state. Simulate the evolution of the particles with Eq.~(\ref{eq-num-algorithm-nondimension}) where $\Vec{j}(\Vec{\lambda})=0$. The driving loop is $\beta\lambda_1= \beta R\cos(2\pi f(s)+\theta_{0})$ and $\beta\lambda_2=\beta R\sin(2\pi f(s)+\theta_{0})$ where $s\equiv t/\tau$, $\theta_0=\pi/6$ and $\beta R=7$. Beginning with a uniform distribution on $x\in[0, L]$, the particles enter the steady state after several driving periods, which means that their reduced probability distribution on $[0,L]$ is periodic. Figures.~\ref{fig: steady-state}(a) and (b) show the evolution of the steady-state distribution under the protocol along path-II and path-III defined in the main text, respectively. The normalized histograms in the upper panels of Figs.~\ref{fig: steady-state}(a) and (b) represent the actual distributions $\rho(x,t)\Delta x$ at different time and those in the lower panels represent the corresponding reduced distributions $\rho_s(x,t)\Delta x\equiv\sum_{n\in\mathbb{Z}}\rho(x+nL,t)\Delta x$. The analytical predictions (red lines in the lower panels) coincide well with the simulation results.

\subsection{Necessity of asymmetric potential}

According to Eq.~(8) of the main text, the reversible integrated flow $\Phi_{\mathrm{rev}}$ reads
\begin{equation}
    \Phi_{\mathrm{rev}}=\oint_I d\Vec{\lambda}\cdot\frac{\int_0^L e^{\beta U_o(x,\Vec{\lambda})}\Vec{f}_o(x,\Vec{\lambda})dx}{\int_0^L e^{\beta U_o(x,\Vec{\lambda})}dx}.
\end{equation}
If the potential $U_o(x,\Vec{\lambda})$ is symmetric, namely, there is always a reference axis to make $U_o(x,\Vec{\lambda})=U_o(-x,\Vec{\lambda})$ for any $x$, then $\rho_o(x,\Vec{\lambda})=\rho_o(-x,\Vec{\lambda})$ and $\Vec{f}_o(x,\Vec{\lambda})=-\Vec{f}_o(-x,\Vec{\lambda})$. Defining $y=-x$, one has
\begin{equation}
\begin{aligned}
    \Phi_{\mathrm{rev}}=&\oint_I d\Vec{\lambda}\cdot\frac{-\int_0^L e^{\beta U_o(-y,\Vec{\lambda})}\Vec{f}_o(-y,\Vec{\lambda})dy}{-\int_0^L e^{\beta U_o(-y,\Vec{\lambda})}dy}\\
    =&\oint_I d\Vec{\lambda}\cdot\frac{-\int_0^L e^{\beta U_o(y,\Vec{\lambda})}\Vec{f}_o(y,\Vec{\lambda})dy}{\int_0^L e^{\beta U_o(y,\Vec{\lambda})}dy}\\
    =&-\Phi_{\mathrm{rev}}\\
    =&0.
\end{aligned}
\end{equation}
Hence, an asymmetric $U_o(x,\Vec{\lambda})$ is needed to generate non-zero $\Phi_{\mathrm{rev}}$ which appears in the first term of the average reduced probability current.

\subsection{Effective force}

The dependence of the effective force on the diffusion coefficient can be used to explain the movements of different particles in the same ratchet potential. The effective force $F_{\mathrm{eff}}$ is defined as the time-ensemble average of force $F(x,t)$, i.e., $F_{\mathrm{eff}}\equiv\tau^{-1}\int_0^{\tau}\int F(x,t)\rho(x,t) dx dt$. For particles driven by $U_o(x,\Vec{\lambda})+U_a^*(x,t)$, $F(x,t)=-\partial_x U_o(x,\Vec{\lambda})-\partial_x U_a^*(x,t)$, then we find
\begin{equation}
    \begin{aligned}
        F_{\mathrm{eff}}=&\frac{1}{\tau}\int_0^{\tau}\int \left[-\frac{\partial U_o(x,\Vec{\lambda})}{\partial x}-\frac{\partial U_a^*(x,t)}{\partial x}\right] \rho(x,t) dxdt\\
        =&\frac{1}{\tau}\int_0^{\tau}\int\frac{1}{\beta} \left(\frac{1}{D}J(x,t)+\frac{\partial\rho(x,t)}{\partial x}\right)dxdt\\
        =&\frac{1}{\tau}\frac{1}{\beta D}\int_0^{\tau}\int J(x,t) dxdt\\
        =&\frac{1}{\tau}\frac{1}{\beta D}\int_0^{\tau}\int_0^L J_s(x,t) dxdt\\
        =&\frac{1}{\beta D}\int_0^L \Bar{J}_s dx\\
        =&\frac{\Bar{v}_s}{\beta D}.
    \end{aligned}
\end{equation}
In the above derivation, we have used the definitions of $J(x,t)$, $J_s(x,t)$, $\Bar{v}_s$ and the fact that $\rho(x,t)$ vanishes at infinity. The direction of $F_{\mathrm{eff}}$ is consistent with the direction of the average velocity which explicitly depends on $D/D^*$ as shown in the main text.

\section{Energetics}

\subsection{Internal energy}
According to the definition of internal energy, we have
\begin{equation}
    \begin{aligned}
        E(t)=&\int U^*(x,t)\rho(x,t)dx\\
        =&\int V^*(x,t)\rho(x,t)dx+\frac{\varepsilon^*(t)}{L}\int x\rho(x,t)dx\\
        =&\sum_{n\in \mathbb{Z}}\int_{nL}^{(n+1)L}V^*(x,t)\rho(x,t)dx+\frac{\varepsilon^*(t)}{L}\langle x \rangle\\
        =&\sum_{n\in \mathbb{Z}}\int_{0}^{L}V^*(x^{\prime}+nL,t)\rho(x^{\prime}+nL,t)dx^{\prime}+\frac{\varepsilon^*(t)}{L}\langle x \rangle\\
        =&\sum_{n\in \mathbb{Z}}\int_{0}^{L}V^*(x^{\prime},t)\rho(x^{\prime}+nL,t)dx^{\prime}+\frac{\varepsilon^*(t)}{L}\langle x \rangle\\
        =&\int_{0}^{L}V^*(x^{\prime},t)\sum_{n\in \mathbb{Z}}\rho(x^{\prime}+nL,t)dx^{\prime}+\frac{\varepsilon^*(t)}{L}\langle x \rangle\\
        =&\int_{0}^{L}V^*(x,t)\rho_s(x,t)dx+\frac{\varepsilon^*(t)}{L}\langle x \rangle,
    \end{aligned}
\end{equation}
where the periodicity of $V^*(x,t)$ and the definition of $\rho_s(x,t)$ have been used. Then we prove that the dependence of $\Delta E$ on the initial time $t_0$ diminishes as the particle transport duration increases. Considering $t_{0}\in[0,\tau]$ is an arbitrary initial time, the variation of the internal energy from $t_{0}$ to $N\tau$ is
\begin{equation}
    \begin{aligned}
        E(N\tau)-E(t_{0})
        =&E(N\tau)-E(0)+E(0)-E(t_{0})\\
        =&N\left[E(\tau)-E(0)\right]+E(0)-E(t_{0})\\
        =&N\varepsilon^*(0)\Bar{J}_s\tau+E(0)-E(t_{0})\\
        =&N\varepsilon^*(0)\Bar{J}_s\tau\left(1+\frac{1}{N}\frac{E(t_0)-E(t^*)}{\varepsilon^*(0)\Bar{J}_s\tau}\right),
    \end{aligned}
\end{equation}
where $E(\tau)-E(0)=\varepsilon^*(0)\Bar{J}_s\tau$ is given by Eq. (12) of the main text. When $N\gg1$, the average energy consumption per single period becomes

\begin{equation}
\frac{E(N\tau)-E(t_0)}{N}\approx\varepsilon^*(0)\Bar{J}_s\tau, 
\end{equation}
which is independent of $t_0$.

\subsection{Ensemble-averaged heat adsorption}

The ensemble-averaged heat current is
\begin{equation}
    \begin{aligned}
        \Dot{Q}=&\int U^*(x,t)\frac{\partial \rho(x,t)}{\partial t}dx\\
        =&-\int U^*(x,t)\frac{\partial J(x,t)}{\partial x}dx\\
        =&-\eval{U^*(x,t)J(x,t)}_{-\infty}^{+\infty}-\int F(x,t)J(x,t)dx\\
        =&-\sum_{n\in \mathbb{Z}}\int_{nL}^{(n+1)L}F(x,t)J(x,t)dx\\
        =&-\sum_{n\in \mathbb{Z}}\int_{0}^{L}F(x+nL,t)J(x+nL,t)dx\\
        =&-\int_0^L F(x,t)J_s(x,t)dx\\
        =&\int_0^L \left[\frac{\partial U_o(x,\Vec{\lambda})}{\partial x}+\frac{\partial U_a^*(x,t)}{\partial x}\right]J_s(x,t)dx,
    \end{aligned}
\end{equation}
where $F(x,t)\equiv-\partial_x U^*(x,t)$ and $J_s(x,t)$ is the reduced probability current. In the above derivation, we have used the fact that $J(x,t)$ vanishes at infinity and the spatial periodicity of $F(x,t)$. Substituting the expressions of $\partial_x U^*(x,t)$ and $J_s(x,t)$ into the above equation, we obtain
\begin{align}
        \Dot{Q}=&\int_0^L\left[\frac{\partial U_o(x,\Vec{\lambda})}{\partial x}+\frac{1}{\beta D^*}\frac{\Dot{\Vec{\lambda}}\cdot\Vec{f}(x,\Vec{\lambda})}{\rho_o(x,\Vec{\lambda})}\right]\left[-\Dot{\Vec{\lambda}}\cdot\Vec{f}(x,\Vec{\lambda})+\left(1-\frac{D}{D^*}\right)\Dot{\Vec{\lambda}}\cdot\langle\Vec{f}(x,\Vec{\lambda})\rangle_+\right]dx  \notag\\
        =&-\int_0^L \frac{\partial U_o(x,\Vec{\lambda})}{\partial x}\Dot{\Vec{\lambda}}\cdot\Vec{f}(x,\Vec{\lambda}) dx+\left(1-\frac{D}{D^*}\right)\Dot{\Vec{\lambda}}\cdot\langle\Vec{f}(x,\Vec{\lambda})\rangle_+\int_0^L \frac{\partial U_o(x,\Vec{\lambda})}{\partial x} dx \notag\\
        &-\frac{1}{\beta D^*}\int_0^L\frac{\left[\Dot{\Vec{\lambda}}\cdot\Vec{f}(x,\Vec{\lambda})\right]^2}{\rho_o(x,\Vec{\lambda})}dx+\frac{1}{\beta D^*}\left(1-\frac{D}{D^*}\right)\Dot{\Vec{\lambda}}\cdot\langle\Vec{f}(x,\Vec{\lambda})\rangle_{+} \int_0^L \frac{\Dot{\Vec{\lambda}}\cdot\Vec{f}(x,\Vec{\lambda})}{\rho_o(x,\Vec{\lambda})} dx \notag\\
        =&-\int_0^L \frac{\partial U_o(x,\Vec{\lambda})}{\partial x}\Dot{\Vec{\lambda}}\cdot\left[ \int_0^x \Vec{\nabla}_{\lambda}\rho_o(x^{\prime},\Vec{\lambda})dx^{\prime}\right] dx-\Dot{\Vec{\lambda}}\cdot\Vec{j}(\Vec{\lambda})\int_0^L \frac{\partial U_o(x,\Vec{\lambda})}{\partial x} dx +0 \notag\\
        &-\frac{1}{\beta D^*} Z_{-}(\Vec{\lambda})Z_{+}(\Vec{\lambda})\left\langle\left[\Dot{\Vec{\lambda}}\cdot\Vec{f}(x,\Vec{\lambda})\right]^2\right\rangle_{+} +\frac{1}{\beta D^*}\left(1-\frac{D}{D^*}\right)Z_{-}(\Vec{\lambda})Z_{+}(\Vec{\lambda})\left[\Dot{\Vec{\lambda}}\cdot\langle\Vec{f}(x,\Vec{\lambda})\rangle_{+} \right]^2 \notag\\
        =&-\eval{\left[ U_o(x,\Vec{\lambda})\Dot{\Vec{\lambda}}\cdot\int_0^x \Vec{\nabla}_{\lambda}\rho_o(x^{\prime},\Vec{\lambda})dx^{\prime}\right]}_{0}^{L} +\int_0^L U_o(x,\Vec{\lambda}) \Dot{\Vec{\lambda}}\cdot\Vec{\nabla}_{\lambda}\rho_o(x,\Vec{\lambda})dx-0 \\
        &-\frac{1}{\beta D^*} Z_{-}(\Vec{\lambda})Z_{+}(\Vec{\lambda})\left\langle\left[\Dot{\Vec{\lambda}}\cdot\Vec{f}(x,\Vec{\lambda})\right]^2\right\rangle_{+} +\frac{1}{\beta D^*}\left(1-\frac{D}{D^*}\right)Z_{-}(\Vec{\lambda})Z_{+}(\Vec{\lambda})\left[\left\langle\Dot{\Vec{\lambda}}\cdot\Vec{f}(x,\Vec{\lambda})\right\rangle_{+} \right]^2 \notag\\
        =&0+\Dot{\Vec{\lambda}}\cdot\Vec{\nabla}_{\lambda} \int_0^L U_o(x,\Vec{\lambda}) \rho_o(x,\Vec{\lambda})dx-\Dot{\Vec{\lambda}}\cdot\int_0^L \left[\Vec{\nabla}_{\lambda}U_o(x,\Vec{\lambda})\right] \rho_o(x,\Vec{\lambda})dx \notag\\
        &-\frac{1}{\beta D^*} Z_{-}(\Vec{\lambda})Z_{+}(\Vec{\lambda})\left\langle\left[\Dot{\Vec{\lambda}}\cdot\Vec{f}(x,\Vec{\lambda})\right]^2\right\rangle_{+} +\frac{1}{\beta D^*}\left(1-\frac{D}{D^*}\right)Z_{-}(\Vec{\lambda})Z_{+}(\Vec{\lambda})\left[\left\langle\Dot{\Vec{\lambda}}\cdot\Vec{f}(x,\Vec{\lambda})\right\rangle_{+} \right]^2 \notag\\
        =&\Dot{\Vec{\lambda}}\cdot\Vec{\nabla}_{\lambda} \int_0^L U_o(x,\Vec{\lambda}) \rho_o(x,\Vec{\lambda})dx+\frac{1}{\beta}\Dot{\Vec{\lambda}}\cdot\Vec{\nabla}_{\lambda}\ln Z_{-} \notag\\
        &-\frac{1}{\beta D^*} Z_{-}(\Vec{\lambda})Z_{+}(\Vec{\lambda})\left\langle\left[\Dot{\Vec{\lambda}}\cdot\Vec{f}(x,\Vec{\lambda})\right]^2\right\rangle_{+} +\frac{1}{\beta D^*}\left(1-\frac{D}{D^*}\right)Z_{-}(\Vec{\lambda})Z_{+}(\Vec{\lambda})\left[\left\langle\Dot{\Vec{\lambda}}\cdot\Vec{f}(x,\Vec{\lambda})\right\rangle_{+} \right]^2 \notag
\end{align}
Then, the heat absorbed by the particle in a temporal period is
\begin{equation}
    \begin{aligned}
        Q=&\int_{t_0}^{t_0+\tau}\Dot{Q}dt\\
        =&\oint_{I}d\Vec{\lambda}\cdot \Vec{\nabla}_{\lambda} \left\{\int_0^L U_o(x,\Vec{\lambda}) \rho_o(x,\Vec{\lambda})dx+\frac{1}{\beta}\ln Z_{-}\right\} \\
        &-\frac{1}{\beta D^*}\int_{t_0}^{t_0+\tau}dt \left\{Z_{-}(\Vec{\lambda})Z_{+}(\Vec{\lambda})\left\langle\left[\Dot{\Vec{\lambda}}\cdot\Vec{f}(x,\Vec{\lambda})\right]^2\right\rangle_{+} - \left(1-\frac{D}{D^*}\right)Z_{-}(\Vec{\lambda})Z_{+}(\Vec{\lambda})\left[\left\langle\Dot{\Vec{\lambda}}\cdot\Vec{f}(x,\Vec{\lambda})\right\rangle_{+} \right]^2 \right\}\\
        =&0-\frac{1}{\beta D^*}\int_{t_0}^{t_0+\tau}dt Z_{-}(\Vec{\lambda})Z_{+}(\Vec{\lambda})\left\{\left\langle\left[\Dot{\Vec{\lambda}}\cdot\Vec{f}(x,\Vec{\lambda})\right]^2\right\rangle_{+} - \left(1-\frac{D}{D^*}\right)\left[\left\langle\Dot{\Vec{\lambda}}\cdot\Vec{f}(x,\Vec{\lambda})\right\rangle_{+} \right]^2 \right\}\\
        =&-\int_{t_0}^{t_0+\tau}dt \Dot{\lambda}_{\alpha}\Dot{\lambda}_{\beta}\frac{1}{\beta D^*}Z_{-}(\Vec{\lambda})Z_{+}(\Vec{\lambda})\left\{\left\langle f_{\alpha}(x,\Vec{\lambda})f_{\beta}(x,\Vec{\lambda})\right\rangle_{+} - \left(1-\frac{D}{D^*}\right)\left\langle f_{\alpha}(x,\Vec{\lambda})\right\rangle_{+}\left\langle f_{\beta}(x,\Vec{\lambda})\right\rangle_{+}  \right\}\\
        \equiv&-\int_{t_0}^{t_0+\tau}dt\Dot{\lambda}_{\alpha}\Dot{\lambda}_{\beta}G_{\alpha\beta}(\Vec{\lambda}),
    \end{aligned}
\end{equation}
where we have used the condition that the  driving trajectory in the parametric space is a loop. 

\subsection{The positive semi-definiteness of $G_{\alpha\beta}(\Vec{\lambda})$}
In the following, We present a proof that $G_{\alpha\beta}(\Vec{\lambda})$ is a positive semi-definite matrix. First, $G_{\alpha\beta}(\Vec{\lambda})=G_{\beta\alpha}(\Vec{\lambda})$ by definition. Secondly, for any $N$-dimensional vector $\Vec{v}=(v_1,v_2,\dots,v_N)$, we have
\begin{equation}
    \begin{aligned}
\Vec{v}G(\Vec{\lambda})\Vec{v}^T=&v_{\alpha}G_{\alpha\beta}(\Vec{\lambda})v_{\beta}\\
        =&v_{\alpha}\frac{1}{\beta D^*}Z_{-}(\Vec{\lambda})Z_{+}(\Vec{\lambda})\left\{\left\langle f_{\alpha}(x,\Vec{\lambda})f_{\beta}(x,\Vec{\lambda})\right\rangle_{+} - \left(1-\frac{D}{D^*}\right)\left\langle f_{\alpha}(x,\Vec{\lambda})\right\rangle_{+}\left\langle f_{\beta}(x,\Vec{\lambda})\right\rangle_{+}  \right\}v_{\beta}\\
        =&\frac{1}{\beta D^*}Z_{-}(\Vec{\lambda})Z_{+}(\Vec{\lambda})\left\{\left\langle  \left[\Vec{v}\cdot\Vec{f}(x,\Vec{\lambda})\right]^2 \right\rangle_{+} - \left(1-\frac{D}{D^*}\right)\left[\left\langle \Vec{v}\cdot\Vec{f}(x,\Vec{\lambda})\right\rangle_{+} \right]^2 \right\}\\
        \ge&\frac{1}{\beta D^*}Z_{-}(\Vec{\lambda})Z_{+}(\Vec{\lambda})\left\{\left[\left\langle  \Vec{v}\cdot\Vec{f}(x,\Vec{\lambda})\right\rangle_{+}\right]^2  - \left(1-\frac{D}{D^*}\right)\left[\left\langle \Vec{v}\cdot\Vec{f}(x,\Vec{\lambda})\right\rangle_{+} \right]^2 \right\}\\
        =&\frac{1}{\beta D^*}Z_{-}(\Vec{\lambda})Z_{+}(\Vec{\lambda})\left\{ \frac{D}{D^*}\left[\left\langle \Vec{v}\cdot\Vec{f}(x,\Vec{\lambda})\right\rangle_{+} \right]^2 \right\}\\
        \ge&0,
    \end{aligned}
\end{equation}
where the inequality $\langle [\Vec{v}\cdot\Vec{f}(x,\Vec{\lambda})]^2 \rangle_{+}\ge [\langle  \Vec{v}\cdot\Vec{f}(x,\Vec{\lambda})\rangle_{+} ]^2$ is obtained via the Cauchy-Schwarz inequality. Here, the inner product of functions $M$ and $N$ is defined as
\begin{equation}
\langle M,N\rangle\equiv\langle MN\rangle_{+}=\frac {\int_0^L e^{\beta U_o(x,\Vec{\lambda})}MNdx}{\int_0^L e^{\beta U_o(x,\Vec{\lambda})}dx}.
\end{equation}
Therefore, $G_{\alpha\beta}$ is positive semi-definite and $Q\le0$.

\section{Simulation details}

\subsection{The dependence of $\Phi_{\mathrm{rev}}$ on $U_o(x,\Vec{\lambda})$ and the driving loop}
\begin{figure}[!tbh]
    \includegraphics{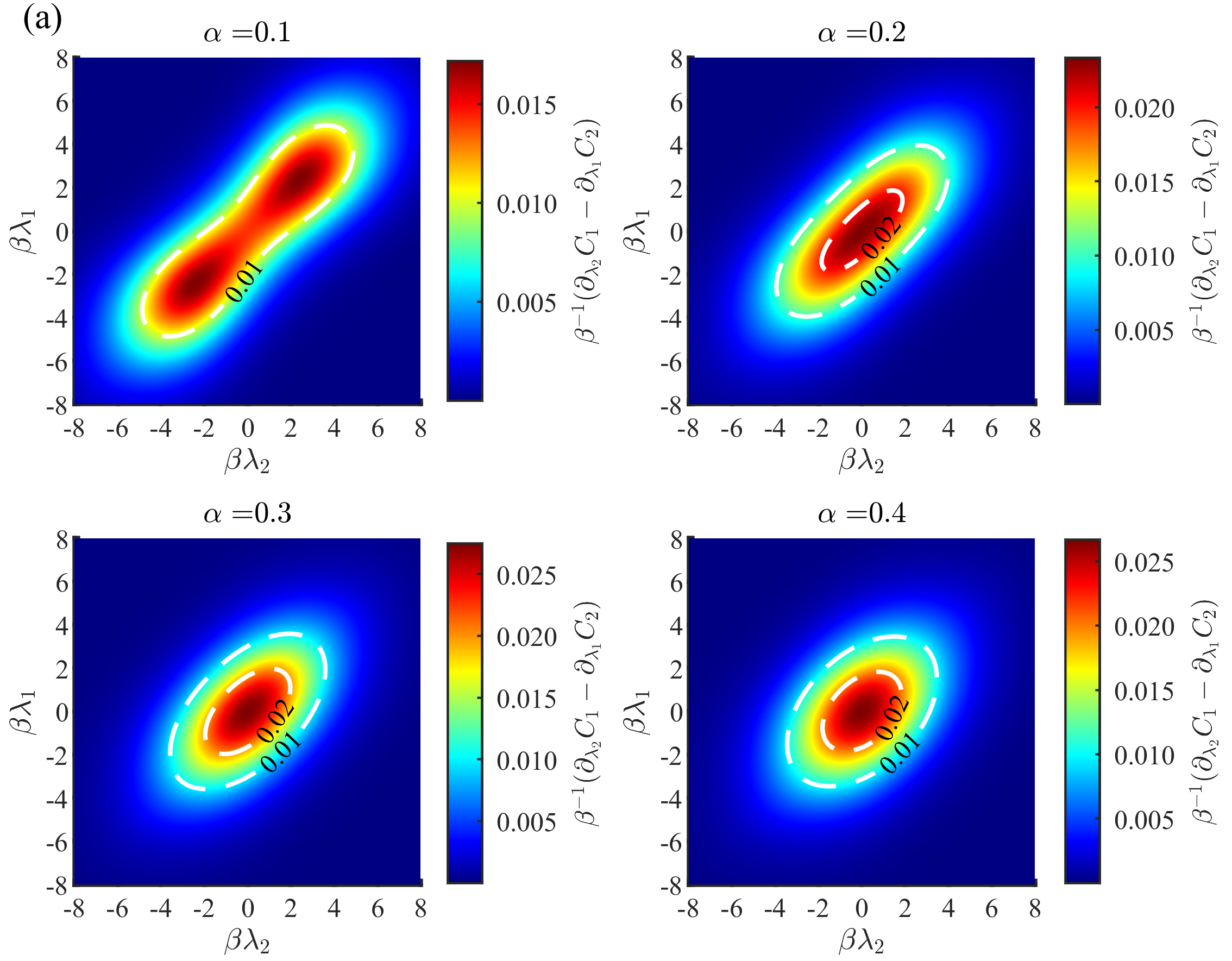}
    \includegraphics{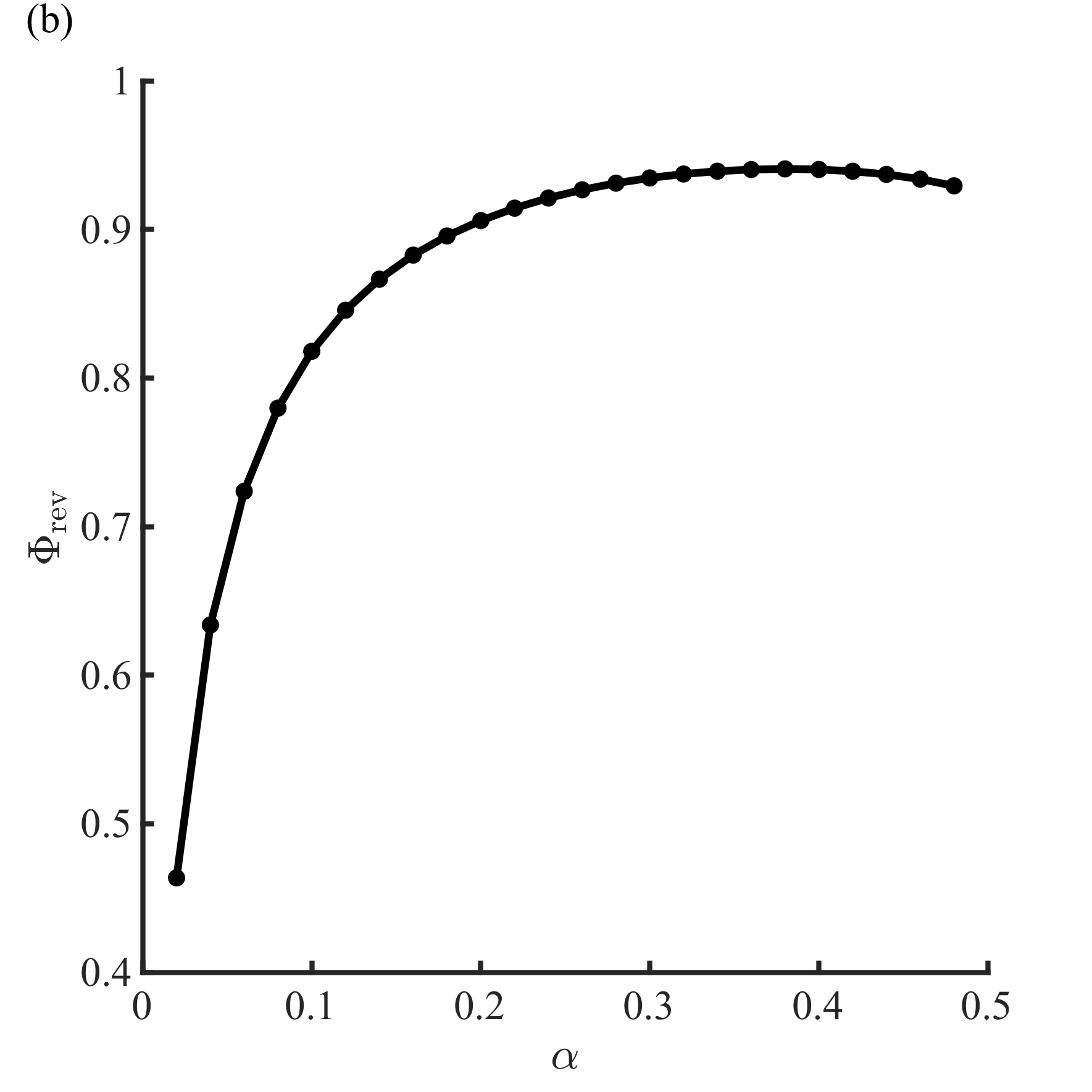}
    \caption{(a) The value of $(\partial_{\lambda_2}C_1-\partial_{\lambda_1}C_2)$ on the parametric space. (b) $\Phi_{\mathrm{rev}}$ as a function of $\alpha$ where the driving loop is a circle with $\beta R$=7}.
    \label{fig: Phirev_optimization}
\end{figure}

The magnitude of the reversible integrated flow $\Phi_{\mathrm{rev}}$ depends on the shape of the potential $U_o(x,\Vec{\lambda})$ and the driving loop in the parametric space. A large $\Phi_{\mathrm{rev}}$ is favorable to induce efficient particle transport. The function $U_o(x,\Vec{\lambda})$ shown in Fig.~2(a) of the main text specifically reads
\begin{equation}
    U_o(x,\Vec{\lambda})=
    \left\{
    \begin{aligned}
        &\frac{\lambda_1}{\alpha}\frac{x}{L},\quad 0\le x/L\le \alpha \\
        &\lambda_1+\frac{\lambda_2-\lambda_1}{1-2\alpha }\left(\frac{x}{L}-\alpha \right),\quad \alpha <x/L\le 1-\alpha \\
        &\lambda_2-\frac{\lambda_2}{\alpha }\left(\frac{x}{L}-1+\alpha \right),\quad 1-\alpha <x/L\le 1,
    \end{aligned}
    \right.
\end{equation}
where $\alpha$ is a tunable parameter. One obtains the expressions of $Z_{\pm}(\Vec{\lambda})$, $\rho_o(x,\Vec{\lambda})$ and $\Vec{f}_o(x,\Vec{\lambda})$ using $U_o(x,\Vec{\lambda})$. According to the Green's theorem, for the two-dimensional parametric vector $\Vec{\lambda}=(\lambda_1,\lambda_2)$, we have
\begin{equation}
    \Phi_{\mathrm{rev}}=\iint_\Sigma\left(\frac{\partial C_1}{\partial \lambda_2}-\frac{\partial C_2}{\partial\lambda_1}\right)d\lambda_1 d\lambda_2.
\end{equation}
Here, $(C_1,C_2)=\Vec{C}(\Vec{\lambda})=-\langle\Vec{f}_o(x,\Vec{\lambda})\rangle_{+}$ is given by Eq.~\eqref{eq-C-expr}, and $\Sigma$ is the region enclosed by $I$. To find favorable $\alpha$ and driving loop, we calculate $(\partial_{\lambda_2}C_1-\partial_{\lambda_1}C_2)$ on the parametric space and $\Phi_{\mathrm{rev}}$ with different $\alpha$, which are shown in Fig.~\ref{fig: Phirev_optimization}. According to Fig.~\ref{fig: Phirev_optimization}(a), $(\partial_{\lambda_2}C_1-\partial_{\lambda_1}C_2)$ is always positive and it exhibits larger values within a closed region containing the origin. It is reasonable to choose the circle with a radius $\beta R$ and a center at the origin as the driving loop $I$. Since $(\partial_{\lambda_2}C_1-\partial_{\lambda_1}C_2)$ is rather small when $\beta\lambda_{1,2}>7$, we set $\beta R=7$. Fig.~\ref{fig: Phirev_optimization}(b) shows $\Phi_{\mathrm{rev}}$ as a function of $\alpha$ with fixed circular driving loop. The maximum value of $\Phi_{\mathrm{rev}}$ appears around $\alpha=0.36$.

\subsection{The optimal driving protocol}

\begin{figure}[!tbh]
    \includegraphics{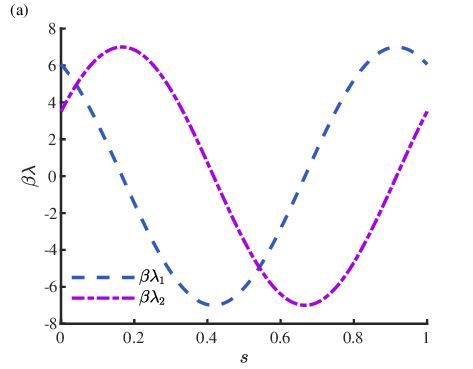}
    \includegraphics{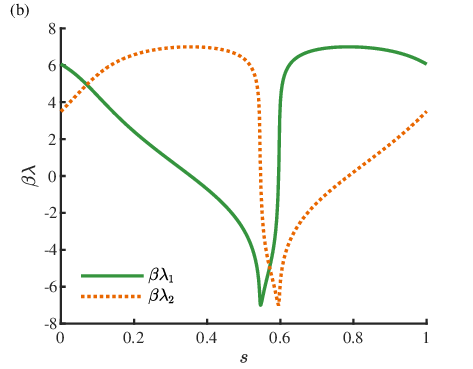}
    \caption{Different driving protocols along the same loop. (a) $f(s)=s$. (b) The optimal protocol.}
    \label{fig: protocols_sin-optimal}
\end{figure}

The optimal protocol $\Vec{\lambda}(t)$ associated with the driving loop $\beta\lambda_1=\beta R\cos(2\pi f(s)+\theta_0)$ and $\beta\lambda_2=\beta R\sin(2\pi f(s)+\theta_0)$ can be numerically obtained from the equation
\begin{equation}
    \Delta s=\frac{\sqrt{\Delta\lambda_{\alpha}\Delta\lambda_{\beta}G_{\alpha\beta}(\Vec{\lambda})}}{\mathcal{L}}\equiv\Delta l,
    \label{eq-optimal-protocol-discrete}
\end{equation}
where $s\equiv t/\tau$. Since the thermodynamic length $\mathcal{L}=\int_0^{\tau}dt\sqrt{\Dot{\lambda}_{\alpha}\Dot{\lambda}_{\beta}G_{\alpha\beta}(\Vec{\lambda})}$ is independent of the protocol \cite{salamonThermodynamicLengthDissipated1983,crooksMeasuringThermodynamicLength2007}, we first calculate the value of $\mathcal{L}$ and generate a series of $\Vec{\lambda}$, namely, $\{\Vec{\lambda}^{(i)}\}$, at $s=0,\Delta,2\Delta,\dots,i\Delta,\dots,n\Delta\equiv 1$ with a simple driving protocol $f(s)=s$ (shown in Fig.~\ref{fig: protocols_sin-optimal}(a)). Here $\Delta$ is a small positive interval, $n\in\mathbb{Z}$ and $0\le i\le n$. Then we calculate the value of $\Delta l^{(j)}$ from $\Vec{\lambda}^{(j-1)}$ to $\Vec{\lambda}^{(j)}$, which gives the optimal variation of $s$ denoted by $\Delta s_{\mathrm{op}}^{(j)}$ through Eq.~(\ref{eq-optimal-protocol-discrete}). Combining the series $\{\Vec{\lambda}^{(i)}\}$ and $\{0,s_{\mathrm{op}}^{(i)}=\sum_{j=1}^{i}\Delta s_{\mathrm{op}}^{(j)}\}$, one obtains the optimal driving protocol illustrated in Fig.~\ref{fig: protocols_sin-optimal}(b).

\subsection{Numerical algorithm for particle dynamics simulation}

We adopt the Euler algorithm of the over-damped Langevin equation to simulate the motion of the Brownian particles of interests~\cite{brankaAlgorithmsBrownianDynamics1998}. The variation of the particle position from $t$ to $t+\Delta t$ is
\begin{equation}
    \Delta x=-\beta D\frac{\partial U^*(x,t)}{\partial x}\Delta t+\sqrt{2D\Delta t}\omega(t),
\end{equation}
where $\omega(t)$ is a standard Gaussian random variable. To reduce the numerical errors in simulations, we perform nondimensionalization on the quantities in the above equation with $L$, $\beta^{-1}$, $D^*$. The nondimensionalized equation reads
\begin{equation}
    \Delta\Tilde{x}=-\Tilde{D}\frac{\partial \Tilde{U}^*(\Tilde{x},\Tilde{t})}{\partial \Tilde{x}}\Delta\Tilde{t}+\sqrt{2\Tilde{D}\Delta\Tilde{t}}\omega(\Tilde{t})\label{eq-num-algorithm-nondimension}
\end{equation}
with $\Tilde{t}=D^*L^{-2}t$, $\Tilde{x}=L^{-1}x$, $\Tilde{D}=D^{*-1}D$, $\Tilde{U}^*=\beta U^*$. In the simulation, the number of particles is $N=10^5$ and the time step is $\Delta\Tilde{t}=10^{-4}$.

\subsection{Heat and work in the simulation}

According to the energetics of the discrete Langevin equation~\cite{sekimotoStochasticEnergetics2010}, the absorbed heat $\Delta q$ and input work $\Delta w$ of the Brownian particles within the time interval $\Delta t$ are given by
\begin{align}
    \Delta q=&U^*(x+\Delta x,t+\Delta t)-U^*(x,t+\Delta t),\label{eq-heat-simulation}\\
    \Delta w=&U^*(x,t+\Delta t)-U^*(x,t),\label{eq-work-simulation}
\end{align}
which are due to the variation of the particle state and the potential parameter, respectively. Obviously, these definitions ensure the energy conservation law $\Delta U^*=\Delta q+\Delta w$ for each particle. In the limit of $\Delta t\to 0$, the ensemble-averaged heat current given by Eq.~\eqref{eq-heat-simulation} becomes
\begin{equation}
\begin{aligned}
\Dot{Q}=&\langle \Dot{q} \rangle\\
       =&\lim_{\Delta t\to 0}\left\langle \frac{U^*(x+\Delta x,t+\Delta t)-U^*(x,t+\Delta t)}{\Delta x}\frac{\Delta x}{\Delta t} \right\rangle \\
       =&\left\langle \frac{\partial U^*(x,t)}{\partial x}\Dot{x} \right\rangle\\
       =&\int \rho (x,t)\frac{\partial U^*(x,t)}{\partial x}\frac{J(x,t)}{\rho(x,t)}dx\\
       =&\eval{U^*(x,t)J(x,t)}_{-\infty}^{+\infty}-\int U^*(x,t)\frac{\partial J(x,t)}{\partial x}dx\\
       =&\int U^*(x,t)\frac{\partial \rho(x,t)}{\partial t},
\end{aligned}
\end{equation}
where $\langle\cdots\rangle$ represents the average over different particle trajectories. This is exactly the same with the definition of heat current we used in the theoretical derivations.

\subsection{Particle separation in 2D space}
We consider a two-dimensional ratchet to separate four kinds of particles respectively with diffusion coefficients $D_1>D_2>D_3>D_4$. We achieve this purpose in two steps: first separate them in the $x$ direction and then divide them into two groups for further separation in the $y$ direction. In the first step, we apply a driving force $F_x(x,t)$ in the $x$ direction to these particles, where $F_x(x,t)=-\partial_x U_o(x,\Vec{\lambda})-\beta^{-1}D_x^{*-1}\rho_o^{-1}(x,\Vec{\lambda})\Dot{\Vec{\lambda}}\cdot\Vec{f}(x,\Vec{\lambda})$ with $D_2>D^*_x>D_3$. The dynamic equations in the first step are
\begin{subequations}
\begin{align}
\Dot{x}=&-\beta D\frac{\partial U_o(x,\Vec{\lambda})}{\partial x}-\frac{D}{D^{*}_x}\frac{\Dot{\Vec{\lambda}}\cdot\Vec{f}_o(x,\Vec{\lambda})}{\rho_o(x,\Vec{\lambda})}+\sqrt{2D}\xi_x(t),\label{eq-2DsepDyn-1st-x}\\
\Dot{y}=&\sqrt{2D}\xi_y(t),\label{eq-2DsepDyn-1st-y}
\end{align}
\end{subequations}
where $\xi_x(t)$ and $\xi_y(t)$ are independent normalized Gaussian white noise. 
The four types of particles will be separated from left to right due to their different average velocities $\Bar{v}_{1x}<\Bar{v}_{2x}<\Bar{v}_{3x}<\Bar{v}_{4x}$.
In the second step when the particles with $D_2$ and $D_3$ have been largely separated, we add the driving force in the $y$ direction on the particles. To achieve efficient separation, the particles with $D_{1,2}$ experience $F_{y\mathrm{L}}(y,t)=-\partial_y U_o(y,\Vec{\lambda})-\beta^{-1}D_{y\mathrm{L}}^{*-1}\rho_o^{-1}(y,\Vec{\lambda})\Dot{\Vec{\lambda}}\cdot\Vec{f}(y,\Vec{\lambda})$ while the particles with $D_{3,4}$ experience $F_{y\mathrm{R}}(y,t)=-\partial_y U_o(y,\Vec{\lambda})-\beta^{-1}D_{y\mathrm{R}}^{*-1}\rho_o^{-1}(y,\Vec{\lambda})\Dot{\Vec{\lambda}}\cdot\Vec{f}(y,\Vec{\lambda})$. Here, $D^*_{y\mathrm{L}}>D^*_{x}>D^*_{y\mathrm{R}}$ The dynamic equations in the second step are
\begin{subequations}
\begin{align}
\Dot{x}=&-\beta D\frac{\partial U_o(x,\Vec{\lambda})}{\partial x}-\frac{D}{D^{*}_x}\frac{\Dot{\Vec{\lambda}}\cdot\Vec{f}_o(x,\Vec{\lambda})}{\rho_o(x,\Vec{\lambda})}+\sqrt{2D}\xi_x(t),\label{eq-2DsepDyn-2nd-x}\\
\Dot{y}=&-\beta D\frac{\partial U_o(y,\Vec{\lambda})}{\partial y}-D\left[\frac{\mathrm{\Theta}(-x+\Bar{v}_{m}t)}{D^{*}_{y\mathrm{L}}}+\frac{\mathrm{\Theta}(x-\Bar{v}_{m}t)}{D^{*}_{y\mathrm{R}}}\right]\frac{\Dot{\Vec{\lambda}}\cdot\Vec{f}_o(y,\Vec{\lambda})}{\rho_o(y,\Vec{\lambda})}+\sqrt{2D}\xi_y(t),\label{eq-2DsepDyn-2nd-y}
\end{align}
\end{subequations}
where $\mathrm{\Theta}(x)$ is the Heaviside step function and $\Bar{v}_{m}\equiv(\Bar{v}_{2x}+\Bar{v}_{3x})/2$. 
The average velocities in the $y$ direction of the particles satisfy $\Bar{v}_{1y}<\Bar{v}_{2y}$ and $\Bar{v}_{3y}<\Bar{v}_{4y}$. The four types of particles will be transported in different directions due to their different velocity vectors.

\begin{figure}
    \centering
    \includegraphics{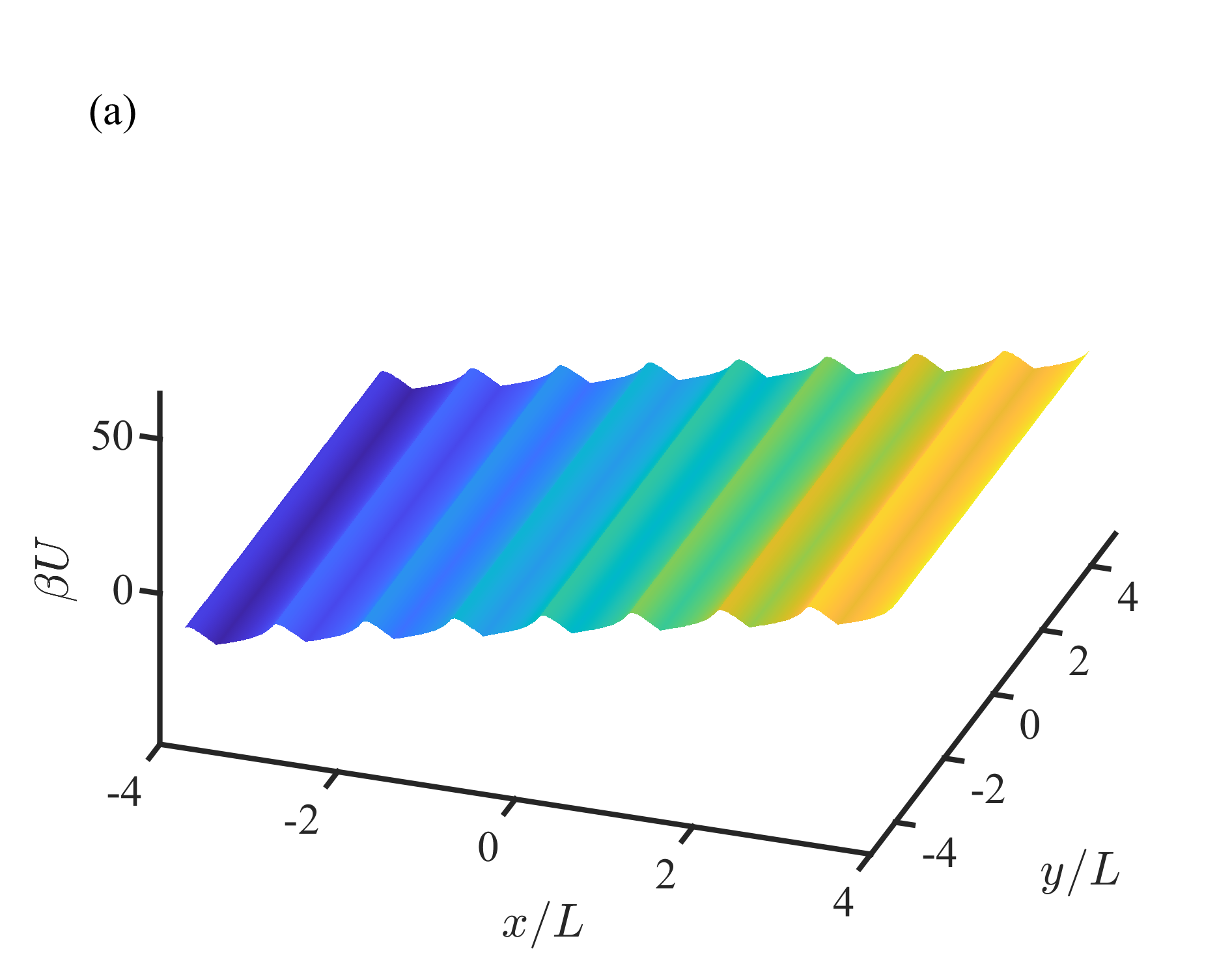}
    \includegraphics{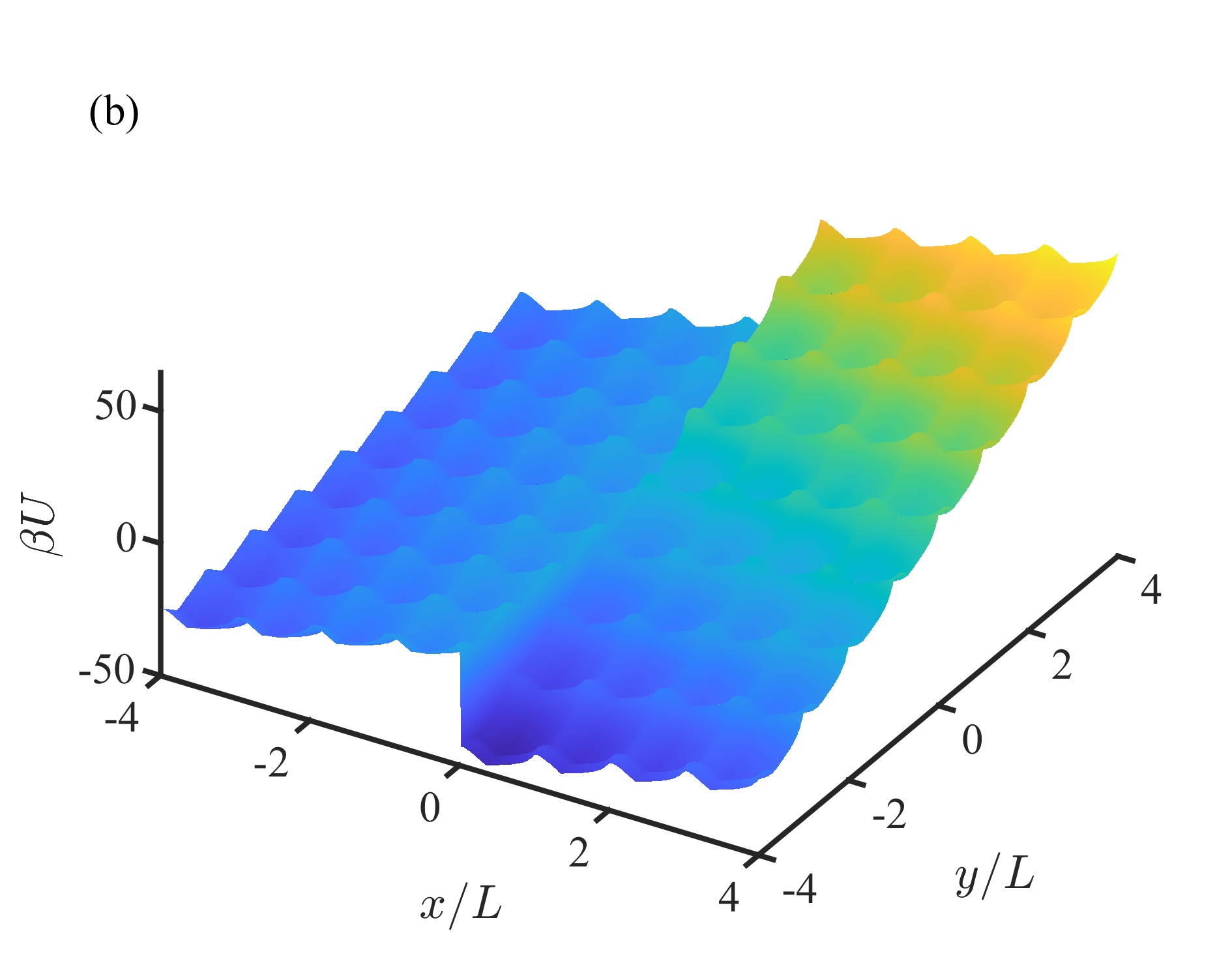}
    \caption{The ratchet potential used for particle separation in 2D space at time $t\mathrm{mod}\tau=0.5\tau$. (a) Potential in the first step. (b) Potential in the second step.}
    \label{fig: 2D-potential}
\end{figure}

We simulate Eqs.~(\ref{eq-2DsepDyn-1st-x}-\ref{eq-2DsepDyn-2nd-y}) with the sawtooth potential and the driving protocol $\beta\lambda_1=7\cos(2\pi t/\tau+\pi/6)$, $\beta\lambda_2=7\sin(2\pi t/\tau+\pi/6)$. $\Vec{j}(\Vec{\lambda})$ is set to be $j_1(\Vec{\lambda})=-\beta^2\lambda_2/30$ and $j_1(\Vec{\lambda})=\beta^2\lambda_1/30$.
The diffusion coefficients in the simulation are $D_1=10D_{m}$, $D_2=25D_{m}/7$, $D_3=5D_{m}/3$, $D_4=D_{m}/5$, $D_x^*=2D_m$, $D_{y\mathrm{L}}^*=25D_{m}/2$ and $D_{y\mathrm{R}}^*=D_{m}$ where $D_{m}$ is a constant with the same dimension of diffusion coefficient used to perform nondimensionalization on the dynamic equations. 
The applied potential is illustrated in Fig.~\ref{fig: 2D-potential}. The driving period is $\tau=10D_{m}^{-1}L^2$, the particles number corresponding to each $D$ is $N=10^5$ and the time step is $\Delta t=5\times 10^{-5}D_{m}^{-1}L^2$. The separation results are shown in Fig.~1(b) of the main text where $\beta=L=D_{m}=1$.

\bibliography{refs_SI}